\documentclass[aps,pra,twocolumn,superscriptaddress]{revtex4-1}  
\usepackage{graphicx,color}  
\usepackage{dcolumn}   
\usepackage{bm}        
\usepackage{amssymb}   
\usepackage{subfigure}
\usepackage{sidecap}
\usepackage{comment}
\usepackage{mathrsfs}
\usepackage{soul}

\providecommand{\ignore}[1]{}

\begin{document}

\title{Anomalous Charge Noise in Superconducting Qubits}

\author{B. G. Christensen} 
\affiliation{Intelligence Community Postdoctoral Research Fellowship Program, University of Wisconsin-Madison, Madison, Wisconsin 53706, USA}

\author{C. D. Wilen} 
\affiliation{Department of Physics, University of Wisconsin-Madison, Madison, Wisconsin 53706, USA}

\author{A. Opremcak} 
\affiliation{Department of Physics, University of Wisconsin-Madison, Madison, Wisconsin 53706, USA}

\author{J. Nelson} 
\affiliation{Department of Physics, Syracuse University, Syracuse, New York 13244, USA}

\author{F. Schlenker} 
\affiliation{Department of Physics, University of Wisconsin-Madison, Madison, Wisconsin 53706, USA}

\author{C. H. Zimonick}
\affiliation{Department of Physics, University of Wisconsin-Madison, Madison, Wisconsin 53706, USA}

\author{L. Faoro} 
\affiliation{Department of Physics, University of Wisconsin-Madison, Madison, Wisconsin 53706, USA}
\affiliation{Sorbonne Universite, Laboratoire de Physique Theorique et Hautes Energies, UMR 7589 CNRS, Tour 13, 5eme Etage, 4 Place Jussieu, F-75252 Paris 05, France}

\author{L. B. Ioffe} 
\affiliation{Department of Physics, University of Wisconsin-Madison, Madison, Wisconsin 53706, USA}
\affiliation{Sorbonne Universite, Laboratoire de Physique Theorique et Hautes Energies, UMR 7589 CNRS, Tour 13, 5eme Etage, 4 Place Jussieu, F-75252 Paris 05, France}

\author{Y. J. Rosen} 
\affiliation{Condensed Matter and Materials Division, Lawrence Livermore National Laboratory, Livermore, California 94550, USA}

\author{J. L. DuBois} 
\affiliation{Condensed Matter and Materials Division, Lawrence Livermore National Laboratory, Livermore, California 94550, USA}

\author{B. L. T. Plourde} 
\affiliation{Department of Physics, Syracuse University, Syracuse, New York 13244, USA}

\author{R. McDermott}
\affiliation{Department of Physics, University of Wisconsin-Madison, Madison, Wisconsin 53706, USA}

\date{\today}

\begin{abstract}
We have used Ramsey tomography to characterize charge noise in a weakly charge-sensitive superconducting qubit. We find a charge noise that scales with frequency as $1/f^\alpha$ over 5 decades with $\alpha = 1.93$ and a magnitude $S_q(\text{1\,Hz})= 2.9\times10^{-4}~e^2/\text{Hz}$. The noise exponent and magnitude of the low-frequency noise are much larger than those seen in prior work on single electron transistors, yet are consistent with reports of frequency noise in other superconducting qubits. Moreover, we observe frequent large-amplitude jumps in offset charge exceeding 0.1$e$; these large discrete charge jumps are incompatible with a picture of localized dipole-like two-level fluctuators. The data reveal an unexpected dependence of charge noise on device scale and suggest models involving either charge drift or fluctuating patch potentials.
\end{abstract}

\maketitle

Superconducting quantum circuits are a leading physical platform for scalable quantum computing, with small-scale qubit arrays nearing the threshold of quantum supremacy \cite{Boixo2018,Neill2018}. The progress of recent years has been enabled by designs that isolate the qubit mode from sources of noise and dissipation inherent in the materials used to realize the device. However, these approaches entail design compromises that could impede continued scaling. For example, the highly successful transmon design \cite{Koch2007} achieves exponential insensitivity against charge noise at the expense of reduced anharmonicity. As a result, leakage out of the computational subspace represents a significant problem for large-scale transmon arrays, as it cannot be mitigated with standard error correction codes \cite{Fowler2013}. At the same time, there are proposals for new qubit designs that provide protection against noise at the hardware level, including charge-parity qubits \cite{Doucot2012,Bell2014}, fluxon pair qubits \cite{Bell2016}, and 0-$\pi$ qubits \cite{Groszkowski2018}. However, in many of these implementations one needs accurate control over the offset charge environment of the device. These considerations motivate a detailed study of charge noise in modern superconducting qubit circuits. 

Previous measurements of charge noise in single electron transistors (SETs) and first-generation charge qubits showed a $1/f$ power spectral density $S_q(f) \propto 1/f^\alpha$ with $\alpha$ between 1.0 and 1.25 \cite{Kuzmin1989, Zimmerli1992, Zimmerli1992M, Visscher1995, Verbrugh1995, Song1995, Wolf1997,Kenyon2000, Nakamura2002, Gustafsson2013, Freeman2016} and noise magnitude $S_q(1\, \rm Hz) \sim 10^{-5}-10^{-7}~e^2/\text{Hz}$. The standard microscopic picture of this noise is a distribution of two-level fluctuators (TLF) \cite{Dutta1981,Kenyon2000,Muller2017} that can activate or tunnel between local minima in a potential energy landscape, leading to switching behavior in the time domain and a Lorentzian power spectral density. A bath of TLF with a broad distribution of characteristic rates gives rise to the ubiquitous $1/f$ noise. 

Here we describe measurements of charge noise in a charge-tunable qubit that departs slightly from the transmon regime. We find a charge noise power spectral density that is up to 4 orders of magnitude larger at 1~Hz than that seen in SETs, suggesting an unexpected dependence of the noise on device scale. Moreover, we observe a large number of discrete charge jumps in excess of $0.1e$. The measured distribution of charge jumps is not compatible with charge motion over microscopic length scales, as described by the standard picture of dipole-like TLF. Finally, the measured noise exponent $\alpha = 1.9$ is incompatible with the exponents reported for SETs, pointing to a new noise mechanism. While our measured noise is strikingly different from that seen in SETs, it is consistent with reports of frequency noise in other superconducting qubits \cite{Riste2013,Serniak2018}.

\begin{figure}[b]
\includegraphics[width=\columnwidth]{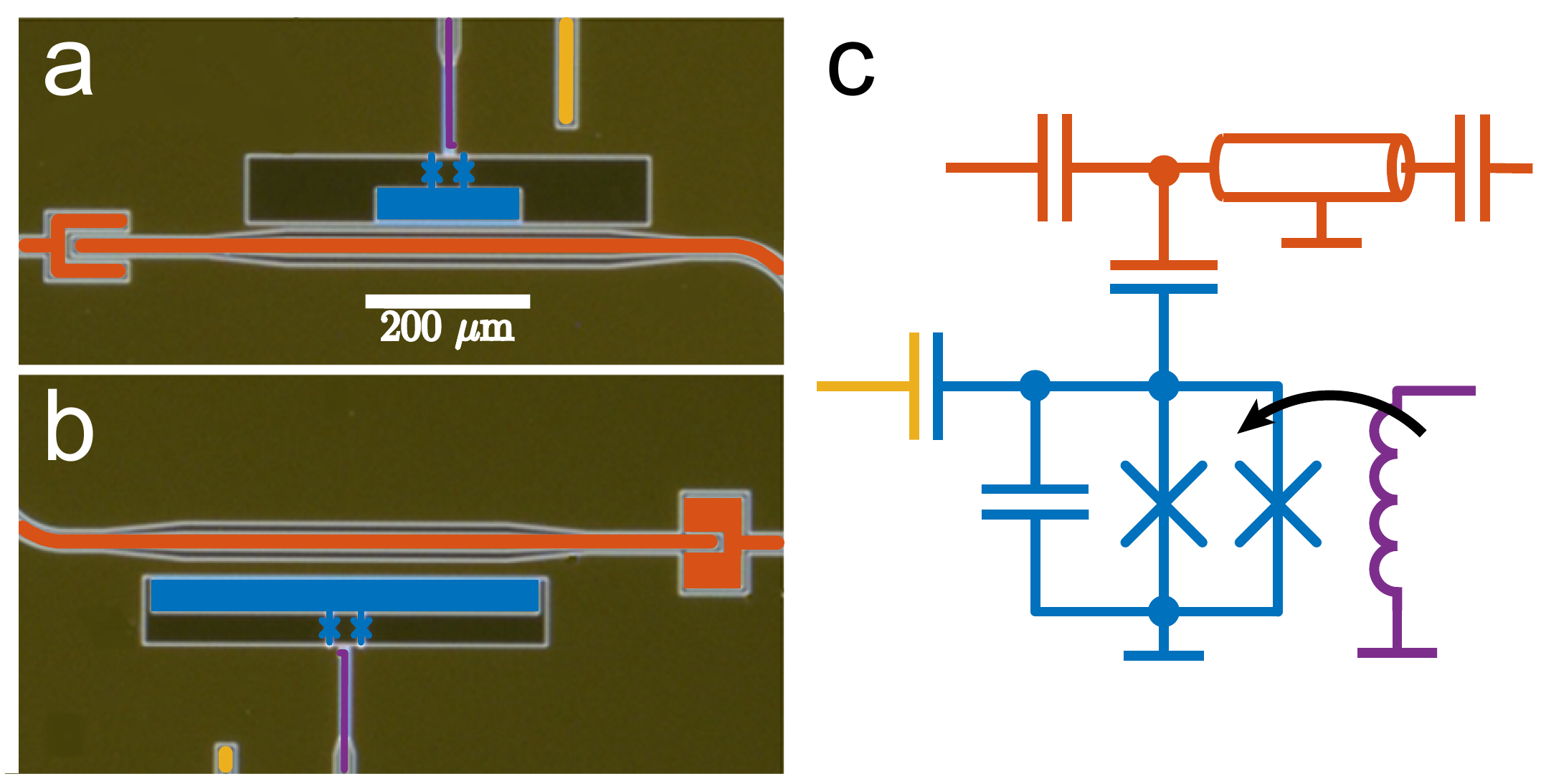}
\caption{\label{fig:device} Micrograph of the charge-sensitive qubit (a; $E_C/h = 390$~MHz, $E_J/h = 10.8$~GHz) and the reference transmon (b; $E_C/h = 230$~MHz, $E_J/h = 16$~GHz).  Here, the qubit structures are shown in blue; the readout resonator and feedline are red; and the charge and flux bias lines are colored orange and purple, respectively. (c) Diagram of the qubit circuit.
}
\end{figure}

\begin{figure*}[t]
\includegraphics[width=\textwidth]{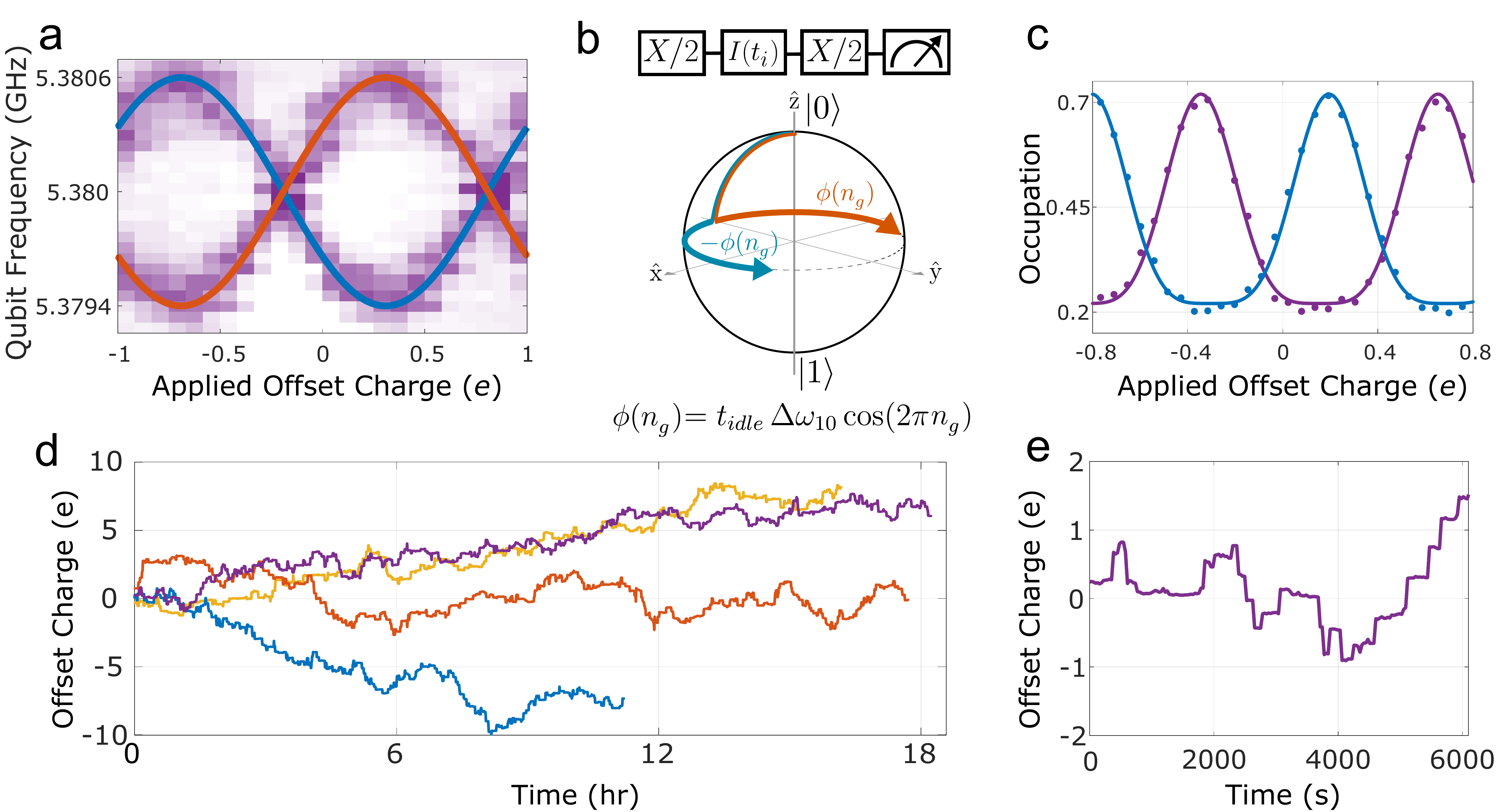}
\caption{\label{fig:spectroscopy} Ramsey-based extraction of offset charge. (a) Qubit spectroscopy versus charge bias at the flux-insensitive point.
As QP tunneling rates far exceed the experimental repetition rate, we observe both QP parity bands (red and blue traces). (b) Pulse sequence used to estimate the offset charge, along with diagram of the trajectory of the qubit state vector on the Bloch sphere for the two values of QP parity.
(c) Representative Ramsey-based charge tomography for two values of $\delta n_g$. The pulse sequence of (b) is repeated for a range of applied offset charge. From a fit to Eq.~(\ref{eq:chargespec}) we extract the change to the island offset charge $\delta n_g$ due to intrinsic noise processes. The full charge sweep is completed in 20~s and the fit uncertainty in $\delta n_g$ is 0.02$e$. (d) Time series of fluctuating offset charge acquired over periods up to 18~hours. (e) Expanded view of measured $\delta n_g$ obtained over a 6000~s interval. Frequent large-magnitude ($> 0.1e$) jumps in offset charge are clearly visible.}
\end{figure*}

The device geometry is shown in Fig.~\ref{fig:device}.  Each die consists of a charge-sensitive qubit and a charge-insensitive reference transmon coupled to a common $\lambda/2$ readout resonator. The devices were fabricated on high-resistivity silicon; the circuit groundplane, qubit islands, and all control and readout elements were made from sputtered niobium and defined using optical lithography and reactive ion etching. The Al-AlO$_x$-Al compound Josephson junctions of the qubits were fabricated using electron-beam lithography and double-angle evaporation. We have performed detailed studies of charge noise in two nominally identical devices. In the following we focus on a single device (qubit A). Data from a second charge-sensitive device (qubit B) are qualitatively similar and are presented in the Supplemental Materials \cite{SuppMat,ChargeFluxNote}.

The parameters for qubit A are $E_J/h = 10.8$~GHz at the flux-insensitive point and $E_C/h = 390$~MHz, corresponding to a qubit transition frequency $\omega_{10}/2\pi =$~5.38~GHz. The readout mode for qubit A resonates at 6.744~GHz. The qubit is coupled to the resonator with a coupling strength of $g/2\pi=$ 100~MHz and the state is read out dispersively with a qubit state-dependent shift of $\chi/\pi = 3.7$ MHz.  The resonator is strongly coupled to the output port with a decay time $1/\kappa = 75$~ns to allow for rapid repeated measurements. The offset charge is controlled through an on-chip capacitance to the qubit island of 100~aF, with a 20:1 voltage division at the millikelvin stage. The device is measured in a dilution refrigerator with a base temperature of 35~mK.

While typical transmon devices involve a ratio $E_J/E_C$ in the range 50-100 \cite{Koch2007, Schreier2008}, leading to a charge dispersion ranging from 10~kHz to 1~Hz, the ratio $E_J/E_C~=~28$ for qubit A yields a charge dispersion $\Delta \omega_{10}/2 \pi$ = 600~kHz. The qubit energy spectrum is given by $\overline{\omega_{10}} + \Delta\omega_{10} \cos(2 \pi n_g)$, where $\overline{\omega_{10}}$ is the charge-averaged qubit frequency and $n_g$ is the offset charge on the qubit island expressed in units of $2e$ (Fig.~\ref{fig:spectroscopy}a). The dependence of the qubit transition frequency on offset charge renders the device sensitive to quasiparticle (QP) poisoning \cite{Lutchyn2006}. Here, single QPs tunnel across the Josephson junctions on sub-millisecond timescales \cite{Riste2013, Serniak2018}, changing $n_g$ by 0.5 and giving rise to distinct parity bands in the qubit spectrum.

\begin{figure}[t!]
\includegraphics[width=\columnwidth]{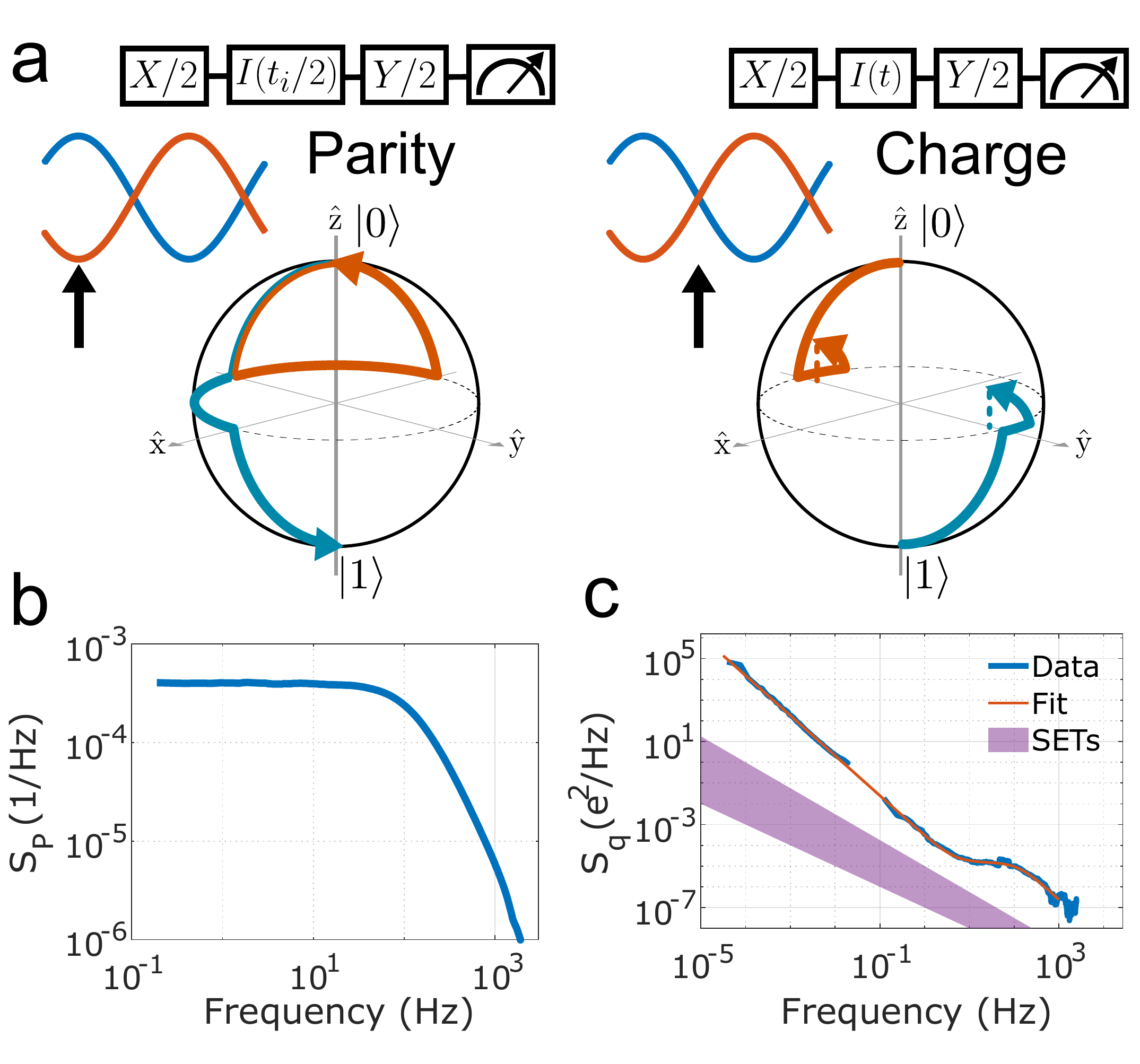}
\caption{\label{fig:pulseseq} (a) Pulse sequences for Ramsey-based single-shot measurement of QP parity (left panel) and charge noise (right panel). In the first sequence, the qubit is biased to a parity-sensitive point; the $X/2$ -- $\textrm{idle}$ -- $Y/2$ sequence is designed to map QP parity to the north and south poles of the Bloch sphere. The sequence is immediately followed by a second experiment that maps fluctuating offset charge to qubit population. Here the device is biased to the point of maximal charge sensitivity; following an initial $X/2$ pulse, states that reside on the two QP parity bands accumulate phase with opposite sign; a final $Y/2$ pulse maps the qubit state to the same polar angle on the Bloch sphere irrespective of QP parity.
(b) Power spectral density of QP parity fluctuations. The spectrum is Lorentzian with a characteristic frequency at $\Gamma/2\pi$~=~255~Hz. (c) Power spectral density of offset charge noise. The low-frequency portion of the spectrum is obtained from the time series presented in Fig.~\ref{fig:spectroscopy}d, while the high-frequency portion of the spectrum is derived from the single-shot protocol of (a). Residual QP tunneling dominates the spectrum above 10~Hz. The orange trace is a fit to the sum of a power law spectrum $S_q(f) = A/f^\alpha$ and a single Lorentzian. We find $S_q(1\,\textrm{Hz}) = 2.9\times10^{-4}~\text{e}^2/\text{Hz}$ and $\alpha = 1.93$.}
\end{figure}

To measure fluctuations in the offset charge on the qubit island, we perform a series of Ramsey experiments at varying charge bias points using a pulse sequence that maps offset charge onto population of the qubit excited state (Fig.~\ref{fig:spectroscopy}b). With QP tunneling rates far exceeding the repetition rate of the Ramsey experiments, we require the experiment to be independent of parity of the qubit island. The sequence begins with a broadband (40~ns long) $X/2$ gate that addresses both parity bands. The qubit then undergoes free evolution for an interval $t_i$, during which time it accumulates the phase $\pm \Delta\omega_{10} t_i \cos 2\pi n_g$, where the two signs correspond to the two possible parity states of the island. While the two parity states evolve in different directions around the equator of the Bloch sphere, they maintain the same projection onto the $y$-axis. We set the idle time $t_i = \pi/\Delta\omega_{10}$ and use a final $X/2$ gate to map this projection onto the $z$-axis of the Bloch sphere.  Measurement of the qubit finds an excited state probability
\begin{equation}
\label{eq:chargespec}
P_1 = \frac{1}{2}\left[d+\nu\cdot\cos(\pi \cos 2\pi n_g)\right],
\end{equation}
where $n_g = n_g^{\rm ext} + \delta n_g$ is the sum of an applied gate charge $n_g^{\rm ext}$ and a fluctuating intrinsic offset charge $\delta n_g$ and the parameters $d$ and $\nu$ account for qubit decay during measurement and finite measurement visibility, respectively. Critically, $P_1$ is periodic in offset charge with period $n_g=0.5$, and is thus insensitive to QP parity.  We sweep the externally applied gate charge $n_g^{\rm ext}$ and determine $\delta n_g$ by fitting the measured Ramsey data to Eq.~(\ref{eq:chargespec}).  Using this technique, we can determine the offset charge to a precision of 0.02$e$ over 20~s. Once $\delta n_g$ is measured, we can then deterministically bias to any point in charge space.

Repeated Ramsey scans of this type generate a time series of fluctuating offset charge; a set of such traces is shown in Fig.~\ref{fig:spectroscopy}d. Interestingly, the charge trace shows occasional (once per $\sim$250~s) extremely large discrete jumps $>0.1e$. The observed distribution of offset charge jumps is difficult to reconcile with a model of dipole-like microscopic TLF; this aspect of the data is discussed in detail below.
Note that, as Ramsey tomography is periodic in an offset charge of 1$e$, we can only determine changes in offset charge within the range $[-0.5e,0.5e)$; any larger jump is aliased to a reduced value of offset charge (e.g., a 0.6$e$ change looks identical to a -0.4$e$ change).

In order to characterize the fluctuating offset charge at higher frequency, we adopt a fast single-shot Ramsey protocol that simultaneously probes island parity and fluctuating offset charge. The measurement sequence is described in Fig.~\ref{fig:pulseseq}a. An initial Ramsey sequence maps the two parity states to the north and south poles of the Bloch sphere. Single-shot measurement of the qubit state provides access to QP parity of the qubit island. Following a short delay of 1 $\mu$s$~\sim 13/\kappa$ to allow the cavity to return to its ground state, we perform a second single-shot Ramsey experiment that maps offset charge to qubit population irrespective of island parity. We bias the qubit to the point of maximal charge sensitivity and perform an $X/2$ gate that rotates the two qubit parity states to opposite sides of the equator of the Bloch sphere. Noise in the charge bias causes the two states to accumulate phase in opposite directions; however, a subsequent $Y/2$ gate maps the accumulated phase to the same polar angle on the Bloch sphere.  Due to the presence of large jumps in offset charge on a few-minute timescale, we interleave with this sequence a separate Ramsey-based scan of offset charge every 15~s in order to characterize and compensate large jumps in offset charge. By repeating the two-step protocol with a duty cycle of 10~kHz, we generate two time series of single-shot measurement results, the first of which provides access to island parity and the second of which provides access to fluctuating offset charge. For each separate time series (QP parity or charge), we partition the time trace into two interleaved traces, compute the cross spectrum, and average over many measurement cycles to suppress quantum projection noise, after \cite{Yan2012, Quintana2017}. 

The measured power spectral densities of QP parity switches and charge noise are shown in Figs.~\ref{fig:pulseseq}b-c. The power spectrum of QP parity is Lorentzian with a characteristic frequency of 255~Hz set by the rate of QP tunneling onto or off of the qubit island; this QP poisoning rate is consistent with other reported values in the superconducting qubit literature \cite{Riste2013,Serniak2018}. For the charge noise results presented in Fig.~\ref{fig:pulseseq}c, we combine the fast single-shot Ramsey results with the low-frequency charge noise power spectral density obtained from the time series presented in Fig.~\ref{fig:spectroscopy}d.
The power spectral density of offset charge fluctuations displays a $1/f^\alpha$ spectrum, with $S_q(1\,\text{Hz}) = 2.9\times10^{-4}~\text{e}^2/\text{Hz}$ and $\alpha = 1.93$. 
The measured charge noise is inconsistent with a large body of literature on charge noise in SETs, both in the noise magnitude at 1~Hz and in the noise exponent. 

While charge noise has not previously been reported on weakly charge-sensitive qubits of the transmon type, there are reports of frequency noise in similar charge-tunable qubits \cite{Riste2013,Serniak2018}. To compare our data to these prior experiments, we convert our measured offset charge to difference frequency using the relation $\delta f = |\Delta\omega_{10} \cos(2 \pi n_g)|$.  In this case, we find $S_{\delta f} (1\,\text{Hz}) = 5.9\times10^{7}~\text{Hz}^2/\text{Hz}$ with noise exponent $\alpha = 1.76$, which closely matches the other measured values (after proper normalization to the same charge dispersion) of $S_{\delta f} (1\,\text{Hz}) = 8.1\times10^{7}~\text{Hz}^2/\text{Hz}$ and $\alpha = 1.7$ \cite{Riste2013}, and $S_{\delta f} (1\,\text{Hz}) = 3.7\times10^{7}~\text{Hz}^2/\text{Hz}$ and $\alpha = 1.70$ \cite{Serniak2018}.  More details on this comparison can be found in the Supplemental Materials \cite{SuppMat}. While a conversion from frequency noise to charge noise is not possible due to non-trivial aliasing effects, the similar levels of frequency noise seen in these three independent qubit measurements suggest a common noise mechanism, despite the fact that these measurements span a range of substrate materials (Si -- this work; Al$_2$O$_3$ -- \cite{Riste2013,Serniak2018}), base metal (Nb -- this work; Al -- \cite{Riste2013,Serniak2018}), and cavity architecture (2D --this work; 3D -- \cite{Riste2013,Serniak2018}).

\begin{figure}[t!]
\includegraphics[width=\columnwidth]{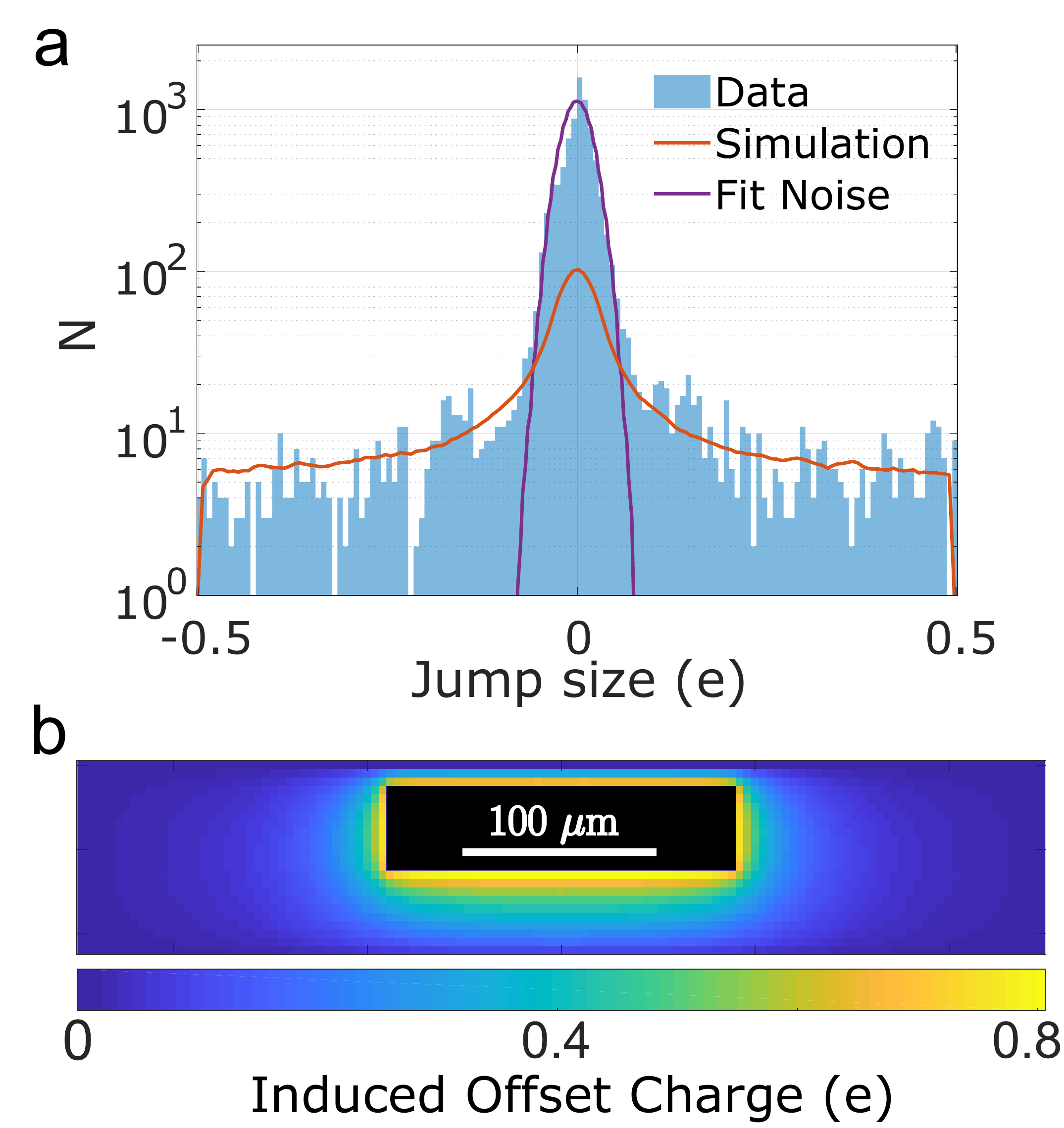}
\caption{\label{fig:hist_and_cap} (a) Histogram of discrete jumps in offset charge taken from 60~hours of data. The histogram displays a large central peak and a long tail of large-magnitude jumps in offset charge. The purple trace is a Gaussian with width 0.02$e$, corresponding to the fit uncertainty in our Ramsey-based charge measurements, while the orange trace is obtained from the numerical simulation in (b), which shows the offset charge associated with impingement of discrete 1$e$ charges in the dielectric space between the qubit island and ground. Here the qubit island is shown in black and field of view extends out to the circuit groundplane. The orange trace in (a) is generated by interpolating the simulation results and aliasing large offset charges to the interval [-0.5$e$, 0.5$e$), as occurs in the measured data.}
\end{figure}

As  this noise is substantially larger than what is seen in SET devices, it is instructive to consider the differences between the two systems. First, SETs are operated in the voltage state, whereas transmons are operated in the superconducting state.  Naively one might expect to observe higher levels of noise in devices operated in the dissipative regime; SET measurements confirm this intuition, where higher voltage bias results in larger noise \cite{Verbrugh1995, Wolf1997}.  The other notable distinction is the large qubit capacitor pad. For our charge-sensitive device, with qubit charging energy $E_C/h = 390$~MHz,
the island dimensions are $40 \times 180$~$\mu$m$^2$. For typical SETs, the island dimensions are submicron and charging energy is of order 40~GHz \cite{Schoelkopf1998,Aassime2001}. It is thus reasonable to consider whether the enhanced noise seen in our devices is related to the difference in device scale. In the most widely accepted picture of low-frequency charge noise, the fluctuating offset charge is due to dipole-like TLF involving motion of a single electron charge over microscopic scales, with dipole moments of order 1~Debye \cite{Martinis2005}. However, such localized dipolar fluctuators are only expected to produce a large change in offset charge when they are located within $\sim$100~nm of the Josephson junctions, which represent the boundary between the island and ground electrodes. However, we observe a broad distribution of discrete jumps in offset charge, with many large jumps in excess of $0.1 e$. In  Fig.~\ref{fig:hist_and_cap} we plot the histogram of discrete charge jumps obtained from the time series in Fig.~\ref{fig:spectroscopy}. In addition to a Gaussian central peak with width 0.02$e$ set by the fit uncertainty in our Ramsey-based charge measurements, the histogram displays long tails corresponding to a large number of discrete charge jumps extending out to $\pm0.5e$ (as described above, any larger charge jumps are aliased into this interval). The frequency of large-magnitude charge jumps suggests a model involving motion or drift of charge as opposed to fluctuations of individual localized TLF. Indeed, the measured histogram is well modeled by random impingement of charge in the dielectric space between the qubit island and the ground electrode (see simulation results in Fig.~\ref{fig:hist_and_cap}a and simulated offset charge in Fig.~\ref{fig:hist_and_cap}b). Moreover, the large noise exponent is consistent with charge drift, as white current noise yields a charge spectrum that scales with frequency as $1/f^2$. For example, it could be that the apparent scale dependence of charge noise is due to a device-dependent sensing area to a fixed background drift of charge in the substrate or in the vacuum environment of the qubit, e.g., due to the motion of ions in the native oxide of the silicon substrate \cite{Gorodokin2004,Constant2000} or to the trapping and release of charged particles in the substrate or surrounding dielectrics due to the relaxation of thermal strain. However, other models are possible, including fluctuating patch potentials on the island electrode \cite{Turchette2000,Deslauriers2004}, for which one would expect the charge noise to scale linearly with the area of the qubit island. We anticipate that systematic study of the dependence of charge noise on device geometry will elucidate the underlying noise mechanism.

In conclusion, we have used a charge-sensitive variant of the transmon qubit to characterize anomalous low-frequency charge noise. The large noise magnitude, the noise exponent approaching 2, and the high density of large discrete charge jumps $>0.1e$ are incompatible with the vast body of literature on charge noise in SETs yet consistent with prior reports of frequency noise in superconducting qubits, indicating a surprising dependence of charge noise on device scale. A deeper understanding of charge noise could guide the development of noise mitigation strategies that will open the design space for superconducting qubits, leading to devices with stronger anharmonicity that are less prone to leakage errors and thus more amenable to scaling.

\smallskip

We acknowledge helpful discussions with Mark Eriksson and Nathan Holman. This research was supported by an appointment to the Intelligence Community Postdoctoral Research Fellowship Program at University of Wisconsin - Madison, administered by Oak Ridge Institute for Science and Education through an interagency agreement between the U.S. Department of Energy and the Office of the Director of National Intelligence. The research is based upon work supported by the Office of the Director of National Intelligence (ODNI), Intelligence Advanced Research Projects Activity (IARPA), via the U.S. Army Research Office grant No. W911NF-16-1-0114.  The views and conclusions contained herein are those of the authors and should not be interpreted as necessarily representing the official policies or endorsements, either expressed or implied, of the ODNI, IARPA, or the U.S. Government. Work by Y.R. and J.D. was performed under the auspices of the U. S. Department of Energy by Lawrence Livermore National Laboratory under Contract No. DE-AC52-07NA27344. Y.R., J.D., B.C. and R. M. acknowledge partial support under LLNL-LDRD SI-16-004.The authors acknowledge use of facilities and instrumentation at the UW-Madison Wisconsin Centers for Nanoscale Technology partially supported by the NSF through the University of Wisconsin Materials Research Science and Engineering Center (DMR-1720415). This work was performed in part at the Cornell NanoScale Science \& Technology Facility (CNF), a member of the National Nanotechnology Coordinated Infrastructure (NNCI), which is supported by the National Science Foundation (Grant NNCI-1542081). LLNL-JRNL-776205.

\bibliographystyle{apsrev4-1-new}
\bibliography{ChargeNoise_BibTeX}

\begin{thebibliography}{39}%
\makeatletter
\providecommand \@ifxundefined [1]{%
 \@ifx{#1\undefined}
}%
\providecommand \@ifnum [1]{%
 \ifnum #1\expandafter \@firstoftwo
 \else \expandafter \@secondoftwo
 \fi
}%
\providecommand \@ifx [1]{%
 \ifx #1\expandafter \@firstoftwo
 \else \expandafter \@secondoftwo
 \fi
}%
\providecommand \natexlab [1]{#1}%
\providecommand \enquote  [1]{``#1''}%
\providecommand \bibnamefont  [1]{#1}%
\providecommand \bibfnamefont [1]{#1}%
\providecommand \citenamefont [1]{#1}%
\providecommand \href@noop [0]{\@secondoftwo}%
\providecommand \href [0]{\begingroup \@sanitize@url \@href}%
\providecommand \@href[1]{\@@startlink{#1}\@@href}%
\providecommand \@@href[1]{\endgroup#1\@@endlink}%
\providecommand \@sanitize@url [0]{\catcode `\\12\catcode `\$12\catcode
  `\&12\catcode `\#12\catcode `\^12\catcode `\_12\catcode `\%12\relax}%
\providecommand \@@startlink[1]{}%
\providecommand \@@endlink[0]{}%
\providecommand \url  [0]{\begingroup\@sanitize@url \@url }%
\providecommand \@url [1]{\endgroup\@href {#1}{\urlprefix }}%
\providecommand \urlprefix  [0]{URL }%
\providecommand \Eprint [0]{\href }%
\providecommand \doibase [0]{http://dx.doi.org/}%
\providecommand \selectlanguage [0]{\@gobble}%
\providecommand \bibinfo  [0]{\@secondoftwo}%
\providecommand \bibfield  [0]{\@secondoftwo}%
\providecommand \translation [1]{[#1]}%
\providecommand \BibitemOpen [0]{}%
\providecommand \bibitemStop [0]{}%
\providecommand \bibitemNoStop [0]{.\EOS\space}%
\providecommand \EOS [0]{\spacefactor3000\relax}%
\providecommand \BibitemShut  [1]{\csname bibitem#1\endcsname}%
\let\auto@bib@innerbib\@empty
\bibitem [{\citenamefont {Boixo}\ \emph {et~al.}(2018)\citenamefont {Boixo},
  \citenamefont {Isakov}, \citenamefont {Smelyanskiy}, \citenamefont {Babbush},
  \citenamefont {Ding}, \citenamefont {Jiang}, \citenamefont {Bremner},
  \citenamefont {Martinis},\ and\ \citenamefont {Neven}}]{Boixo2018}%
  \BibitemOpen
  \bibfield  {author} {\bibinfo {author} {\bibfnamefont {S.}~\bibnamefont
  {Boixo}}, \bibinfo {author} {\bibfnamefont {S.~V.}\ \bibnamefont {Isakov}},
  \bibinfo {author} {\bibfnamefont {V.~N.}\ \bibnamefont {Smelyanskiy}},
  \bibinfo {author} {\bibfnamefont {R.}~\bibnamefont {Babbush}}, \bibinfo
  {author} {\bibfnamefont {N.}~\bibnamefont {Ding}}, \bibinfo {author}
  {\bibfnamefont {Z.}~\bibnamefont {Jiang}}, \bibinfo {author} {\bibfnamefont
  {M.~J.}\ \bibnamefont {Bremner}}, \bibinfo {author} {\bibfnamefont {J.~M.}\
  \bibnamefont {Martinis}}, \ and\ \bibinfo {author} {\bibfnamefont
  {H.}~\bibnamefont {Neven}},\ }\bibfield  {title} {\emph {\bibinfo {title}
  {Characterizing quantum supremacy in near-term devices},\ }}\href
  {https://doi.org/10.1038/s41567-018-0124-x} {\bibfield  {journal} {\bibinfo
  {journal} {Nature Physics}\ }\textbf {\bibinfo {volume} {14}},\ \bibinfo
  {pages} {595} (\bibinfo {year} {2018})}\BibitemShut {NoStop}%
\bibitem [{\citenamefont {Neill}\ \emph {et~al.}(2018)\citenamefont {Neill},
  \citenamefont {Roushan}, \citenamefont {Kechedzhi}, \citenamefont {Boixo},
  \citenamefont {Isakov}, \citenamefont {Smelyanskiy}, \citenamefont {Megrant},
  \citenamefont {Chiaro}, \citenamefont {Dunsworth}, \citenamefont {Arya},
  \citenamefont {Barends}, \citenamefont {Burkett}, \citenamefont {Chen},
  \citenamefont {Chen}, \citenamefont {Fowler}, \citenamefont {Foxen},
  \citenamefont {Giustina}, \citenamefont {Graff}, \citenamefont {Jeffrey},
  \citenamefont {Huang}, \citenamefont {Kelly}, \citenamefont {Klimov},
  \citenamefont {Lucero}, \citenamefont {Mutus}, \citenamefont {Neeley},
  \citenamefont {Quintana}, \citenamefont {Sank}, \citenamefont {Vainsencher},
  \citenamefont {Wenner}, \citenamefont {White}, \citenamefont {Neven},\ and\
  \citenamefont {Martinis}}]{Neill2018}%
  \BibitemOpen
  \bibfield  {author} {\bibinfo {author} {\bibfnamefont {C.}~\bibnamefont
  {Neill}}, \bibinfo {author} {\bibfnamefont {P.}~\bibnamefont {Roushan}},
  \bibinfo {author} {\bibfnamefont {K.}~\bibnamefont {Kechedzhi}}, \bibinfo
  {author} {\bibfnamefont {S.}~\bibnamefont {Boixo}}, \bibinfo {author}
  {\bibfnamefont {S.~V.}\ \bibnamefont {Isakov}}, \bibinfo {author}
  {\bibfnamefont {V.}~\bibnamefont {Smelyanskiy}}, \bibinfo {author}
  {\bibfnamefont {A.}~\bibnamefont {Megrant}}, \bibinfo {author} {\bibfnamefont
  {B.}~\bibnamefont {Chiaro}}, \bibinfo {author} {\bibfnamefont
  {A.}~\bibnamefont {Dunsworth}}, \bibinfo {author} {\bibfnamefont
  {K.}~\bibnamefont {Arya}}, \bibinfo {author} {\bibfnamefont {R.}~\bibnamefont
  {Barends}}, \bibinfo {author} {\bibfnamefont {B.}~\bibnamefont {Burkett}},
  \bibinfo {author} {\bibfnamefont {Y.}~\bibnamefont {Chen}}, \bibinfo {author}
  {\bibfnamefont {Z.}~\bibnamefont {Chen}}, \bibinfo {author} {\bibfnamefont
  {A.}~\bibnamefont {Fowler}}, \bibinfo {author} {\bibfnamefont
  {B.}~\bibnamefont {Foxen}}, \bibinfo {author} {\bibfnamefont
  {M.}~\bibnamefont {Giustina}}, \bibinfo {author} {\bibfnamefont
  {R.}~\bibnamefont {Graff}}, \bibinfo {author} {\bibfnamefont
  {E.}~\bibnamefont {Jeffrey}}, \bibinfo {author} {\bibfnamefont
  {T.}~\bibnamefont {Huang}}, \bibinfo {author} {\bibfnamefont
  {J.}~\bibnamefont {Kelly}}, \bibinfo {author} {\bibfnamefont
  {P.}~\bibnamefont {Klimov}}, \bibinfo {author} {\bibfnamefont
  {E.}~\bibnamefont {Lucero}}, \bibinfo {author} {\bibfnamefont
  {J.}~\bibnamefont {Mutus}}, \bibinfo {author} {\bibfnamefont
  {M.}~\bibnamefont {Neeley}}, \bibinfo {author} {\bibfnamefont
  {C.}~\bibnamefont {Quintana}}, \bibinfo {author} {\bibfnamefont
  {D.}~\bibnamefont {Sank}}, \bibinfo {author} {\bibfnamefont {A.}~\bibnamefont
  {Vainsencher}}, \bibinfo {author} {\bibfnamefont {J.}~\bibnamefont {Wenner}},
  \bibinfo {author} {\bibfnamefont {T.~C.}\ \bibnamefont {White}}, \bibinfo
  {author} {\bibfnamefont {H.}~\bibnamefont {Neven}}, \ and\ \bibinfo {author}
  {\bibfnamefont {J.~M.}\ \bibnamefont {Martinis}},\ }\bibfield  {title} {\emph
  {\bibinfo {title} {A blueprint for demonstrating quantum supremacy with
  superconducting qubits},\ }}\href {\doibase 10.1126/science.aao4309}
  {\bibfield  {journal} {\bibinfo  {journal} {Science}\ }\textbf {\bibinfo
  {volume} {360}},\ \bibinfo {pages} {195} (\bibinfo {year} {2018})},\ \Eprint
  {http://arxiv.org/abs/http://science.sciencemag.org/content/360/6385/195.full.pdf}
  {http://science.sciencemag.org/content/360/6385/195.full.pdf} \BibitemShut
  {NoStop}%
\bibitem [{\citenamefont {Koch}\ \emph {et~al.}(2007)\citenamefont {Koch},
  \citenamefont {Yu}, \citenamefont {Gambetta}, \citenamefont {Houck},
  \citenamefont {Schuster}, \citenamefont {Majer}, \citenamefont {Blais},
  \citenamefont {Devoret}, \citenamefont {Girvin},\ and\ \citenamefont
  {Schoelkopf}}]{Koch2007}%
  \BibitemOpen
  \bibfield  {author} {\bibinfo {author} {\bibfnamefont {J.}~\bibnamefont
  {Koch}}, \bibinfo {author} {\bibfnamefont {T.~M.}\ \bibnamefont {Yu}},
  \bibinfo {author} {\bibfnamefont {J.}~\bibnamefont {Gambetta}}, \bibinfo
  {author} {\bibfnamefont {A.~A.}\ \bibnamefont {Houck}}, \bibinfo {author}
  {\bibfnamefont {D.~I.}\ \bibnamefont {Schuster}}, \bibinfo {author}
  {\bibfnamefont {J.}~\bibnamefont {Majer}}, \bibinfo {author} {\bibfnamefont
  {A.}~\bibnamefont {Blais}}, \bibinfo {author} {\bibfnamefont {M.~H.}\
  \bibnamefont {Devoret}}, \bibinfo {author} {\bibfnamefont {S.~M.}\
  \bibnamefont {Girvin}}, \ and\ \bibinfo {author} {\bibfnamefont {R.~J.}\
  \bibnamefont {Schoelkopf}},\ }\bibfield  {title} {\emph {\bibinfo {title}
  {Charge-insensitive qubit design derived from the cooper pair box},\ }}\href
  {\doibase 10.1103/PhysRevA.76.042319} {\bibfield  {journal} {\bibinfo
  {journal} {Phys. Rev. A}\ }\textbf {\bibinfo {volume} {76}},\ \bibinfo
  {pages} {042319} (\bibinfo {year} {2007})}\BibitemShut {NoStop}%
\bibitem [{\citenamefont {Fowler}(2013)}]{Fowler2013}%
  \BibitemOpen
  \bibfield  {author} {\bibinfo {author} {\bibfnamefont {A.~G.}\ \bibnamefont
  {Fowler}},\ }\bibfield  {title} {\emph {\bibinfo {title} {Coping with qubit
  leakage in topological codes},\ }}\href
  {https://link.aps.org/doi/10.1103/PhysRevA.88.042308} {\bibfield  {journal}
  {\bibinfo  {journal} {Phys. Rev. A}\ }\textbf {\bibinfo {volume} {88}},\
  \bibinfo {pages} {042308} (\bibinfo {year} {2013})}\BibitemShut {NoStop}%
\bibitem [{\citenamefont {Dou\c{c}ot}\ and\ \citenamefont
  {Ioffe}(2012)}]{Doucot2012}%
  \BibitemOpen
  \bibfield  {author} {\bibinfo {author} {\bibfnamefont {B.}~\bibnamefont
  {Dou\c{c}ot}}\ and\ \bibinfo {author} {\bibfnamefont {L.~B.}\ \bibnamefont
  {Ioffe}},\ }\bibfield  {title} {\emph {\bibinfo {title} {Physical
  implementation of protected qubits},\ }}\href
  {http://dx.doi.org/10.1088/0034-4885/75/7/072001} {\bibfield  {journal}
  {\bibinfo  {journal} {Reports on Progress in Physics}\ }\textbf {\bibinfo
  {volume} {75}},\ \bibinfo {pages} {072001} (\bibinfo {year}
  {2012})}\BibitemShut {NoStop}%
\bibitem [{\citenamefont {Bell}\ \emph {et~al.}(2014)\citenamefont {Bell},
  \citenamefont {Paramanandam}, \citenamefont {Ioffe},\ and\ \citenamefont
  {Gershenson}}]{Bell2014}%
  \BibitemOpen
  \bibfield  {author} {\bibinfo {author} {\bibfnamefont {M.~T.}\ \bibnamefont
  {Bell}}, \bibinfo {author} {\bibfnamefont {J.}~\bibnamefont {Paramanandam}},
  \bibinfo {author} {\bibfnamefont {L.~B.}\ \bibnamefont {Ioffe}}, \ and\
  \bibinfo {author} {\bibfnamefont {M.~E.}\ \bibnamefont {Gershenson}},\
  }\bibfield  {title} {\emph {\bibinfo {title} {Protected josephson rhombus
  chains},\ }}\href {https://link.aps.org/doi/10.1103/PhysRevLett.112.167001}
  {\bibfield  {journal} {\bibinfo  {journal} {Phys. Rev. Lett.}\ }\textbf
  {\bibinfo {volume} {112}},\ \bibinfo {pages} {167001} (\bibinfo {year}
  {2014})}\BibitemShut {NoStop}%
\bibitem [{\citenamefont {Bell}\ \emph {et~al.}(2016)\citenamefont {Bell},
  \citenamefont {Zhang}, \citenamefont {Ioffe},\ and\ \citenamefont
  {Gershenson}}]{Bell2016}%
  \BibitemOpen
  \bibfield  {author} {\bibinfo {author} {\bibfnamefont {M.~T.}\ \bibnamefont
  {Bell}}, \bibinfo {author} {\bibfnamefont {W.}~\bibnamefont {Zhang}},
  \bibinfo {author} {\bibfnamefont {L.~B.}\ \bibnamefont {Ioffe}}, \ and\
  \bibinfo {author} {\bibfnamefont {M.~E.}\ \bibnamefont {Gershenson}},\
  }\bibfield  {title} {\emph {\bibinfo {title} {Spectroscopic evidence of the
  {A}haronov-{C}asher effect in a cooper pair box},\ }}\href
  {https://link.aps.org/doi/10.1103/PhysRevLett.116.107002} {\bibfield
  {journal} {\bibinfo  {journal} {Phys. Rev. Lett.}\ }\textbf {\bibinfo
  {volume} {116}},\ \bibinfo {pages} {107002} (\bibinfo {year}
  {2016})}\BibitemShut {NoStop}%
\bibitem [{\citenamefont {Groszkowski}\ \emph {et~al.}(2018)\citenamefont
  {Groszkowski}, \citenamefont {Paolo}, \citenamefont {Grimsmo}, \citenamefont
  {Blais}, \citenamefont {Schuster}, \citenamefont {Houck},\ and\ \citenamefont
  {Koch}}]{Groszkowski2018}%
  \BibitemOpen
  \bibfield  {author} {\bibinfo {author} {\bibfnamefont {P.}~\bibnamefont
  {Groszkowski}}, \bibinfo {author} {\bibfnamefont {A.~D.}\ \bibnamefont
  {Paolo}}, \bibinfo {author} {\bibfnamefont {A.~L.}\ \bibnamefont {Grimsmo}},
  \bibinfo {author} {\bibfnamefont {A.}~\bibnamefont {Blais}}, \bibinfo
  {author} {\bibfnamefont {D.~I.}\ \bibnamefont {Schuster}}, \bibinfo {author}
  {\bibfnamefont {A.~A.}\ \bibnamefont {Houck}}, \ and\ \bibinfo {author}
  {\bibfnamefont {J.}~\bibnamefont {Koch}},\ }\bibfield  {title} {\emph
  {\bibinfo {title} {Coherence properties of the 0-pi qubit},\ }}\href
  {http://dx.doi.org/10.1088/1367-2630/aab7cd} {\bibfield  {journal} {\bibinfo
  {journal} {New Journal of Physics}\ }\textbf {\bibinfo {volume} {20}},\
  \bibinfo {pages} {043053} (\bibinfo {year} {2018})}\BibitemShut {NoStop}%
\bibitem [{\citenamefont {Kuzmin}\ \emph {et~al.}(1989)\citenamefont {Kuzmin},
  \citenamefont {Delsing}, \citenamefont {Claeson},\ and\ \citenamefont
  {Likharev}}]{Kuzmin1989}%
  \BibitemOpen
  \bibfield  {author} {\bibinfo {author} {\bibfnamefont {L.~S.}\ \bibnamefont
  {Kuzmin}}, \bibinfo {author} {\bibfnamefont {P.}~\bibnamefont {Delsing}},
  \bibinfo {author} {\bibfnamefont {T.}~\bibnamefont {Claeson}}, \ and\
  \bibinfo {author} {\bibfnamefont {K.~K.}\ \bibnamefont {Likharev}},\
  }\bibfield  {title} {\emph {\bibinfo {title} {Single-electron charging
  effects in one-dimensional arrays of ultrasmall tunnel junctions},\ }}\href
  {https://link.aps.org/doi/10.1103/PhysRevLett.62.2539} {\bibfield  {journal}
  {\bibinfo  {journal} {Phys. Rev. Lett.}\ }\textbf {\bibinfo {volume} {62}},\
  \bibinfo {pages} {2539} (\bibinfo {year} {1989})}\BibitemShut {NoStop}%
\bibitem [{\citenamefont {Zimmerli}\ \emph
  {et~al.}(1992{\natexlab{a}})\citenamefont {Zimmerli}, \citenamefont {Eiles},
  \citenamefont {Kautz},\ and\ \citenamefont {Martinis}}]{Zimmerli1992}%
  \BibitemOpen
  \bibfield  {author} {\bibinfo {author} {\bibfnamefont {G.}~\bibnamefont
  {Zimmerli}}, \bibinfo {author} {\bibfnamefont {T.~M.}\ \bibnamefont {Eiles}},
  \bibinfo {author} {\bibfnamefont {R.~L.}\ \bibnamefont {Kautz}}, \ and\
  \bibinfo {author} {\bibfnamefont {J.~M.}\ \bibnamefont {Martinis}},\
  }\bibfield  {title} {\emph {\bibinfo {title} {Noise in the coulomb blockade
  electrometer},\ }}\href {\doibase 10.1063/1.108195} {\bibfield  {journal}
  {\bibinfo  {journal} {Appl. Phys. Lett.}\ }\textbf {\bibinfo {volume} {61}},\
  \bibinfo {pages} {237} (\bibinfo {year} {1992}{\natexlab{a}})}\BibitemShut
  {NoStop}%
\bibitem [{\citenamefont {Zimmerli}\ \emph
  {et~al.}(1992{\natexlab{b}})\citenamefont {Zimmerli}, \citenamefont {Kautz},\
  and\ \citenamefont {Martinis}}]{Zimmerli1992M}%
  \BibitemOpen
  \bibfield  {author} {\bibinfo {author} {\bibfnamefont {G.}~\bibnamefont
  {Zimmerli}}, \bibinfo {author} {\bibfnamefont {R.~L.}\ \bibnamefont {Kautz}},
  \ and\ \bibinfo {author} {\bibfnamefont {J.~M.}\ \bibnamefont {Martinis}},\
  }\bibfield  {title} {\emph {\bibinfo {title} {Voltage gain in the
  single-electron transistor},\ }}\href {\doibase 10.1063/1.108117} {\bibfield
  {journal} {\bibinfo  {journal} {Appl. Phys. Lett.}\ }\textbf {\bibinfo
  {volume} {61}},\ \bibinfo {pages} {2616} (\bibinfo {year}
  {1992}{\natexlab{b}})}\BibitemShut {NoStop}%
\bibitem [{\citenamefont {Visscher}\ \emph {et~al.}(1995)\citenamefont
  {Visscher}, \citenamefont {Verbrugh}, \citenamefont {Lindeman}, \citenamefont
  {Hadley},\ and\ \citenamefont {Mooij}}]{Visscher1995}%
  \BibitemOpen
  \bibfield  {author} {\bibinfo {author} {\bibfnamefont {E.~H.}\ \bibnamefont
  {Visscher}}, \bibinfo {author} {\bibfnamefont {S.~M.}\ \bibnamefont
  {Verbrugh}}, \bibinfo {author} {\bibfnamefont {J.}~\bibnamefont {Lindeman}},
  \bibinfo {author} {\bibfnamefont {P.}~\bibnamefont {Hadley}}, \ and\ \bibinfo
  {author} {\bibfnamefont {J.~E.}\ \bibnamefont {Mooij}},\ }\bibfield  {title}
  {\emph {\bibinfo {title} {Fabrication of multilayer single-electron tunneling
  devices},\ }}\href {\doibase 10.1063/1.113526} {\bibfield  {journal}
  {\bibinfo  {journal} {Appl. Phys. Lett.}\ }\textbf {\bibinfo {volume} {66}},\
  \bibinfo {pages} {305} (\bibinfo {year} {1995})}\BibitemShut {NoStop}%
\bibitem [{\citenamefont {Verbrugh}\ \emph {et~al.}(1995)\citenamefont
  {Verbrugh}, \citenamefont {Benhamadi}, \citenamefont {Visscher},\ and\
  \citenamefont {Mooij}}]{Verbrugh1995}%
  \BibitemOpen
  \bibfield  {author} {\bibinfo {author} {\bibfnamefont {S.~M.}\ \bibnamefont
  {Verbrugh}}, \bibinfo {author} {\bibfnamefont {M.~L.}\ \bibnamefont
  {Benhamadi}}, \bibinfo {author} {\bibfnamefont {E.~H.}\ \bibnamefont
  {Visscher}}, \ and\ \bibinfo {author} {\bibfnamefont {J.~E.}\ \bibnamefont
  {Mooij}},\ }\bibfield  {title} {\emph {\bibinfo {title} {Optimization of
  island size in single electron tunneling devices: Experiment and theory},\
  }}\href {\doibase 10.1063/1.360083} {\bibfield  {journal} {\bibinfo
  {journal} {Journal of Applied Physics}\ }\textbf {\bibinfo {volume} {78}},\
  \bibinfo {pages} {2830} (\bibinfo {year} {1995})}\BibitemShut {NoStop}%
\bibitem [{\citenamefont {Song}\ \emph {et~al.}(1995)\citenamefont {Song},
  \citenamefont {Amar}, \citenamefont {Lobb},\ and\ \citenamefont
  {Wellstood}}]{Song1995}%
  \BibitemOpen
  \bibfield  {author} {\bibinfo {author} {\bibfnamefont {D.}~\bibnamefont
  {Song}}, \bibinfo {author} {\bibfnamefont {A.}~\bibnamefont {Amar}}, \bibinfo
  {author} {\bibfnamefont {C.~J.}\ \bibnamefont {Lobb}}, \ and\ \bibinfo
  {author} {\bibfnamefont {F.~C.}\ \bibnamefont {Wellstood}},\ }\bibfield
  {title} {\emph {\bibinfo {title} {Advantages of superconducting
  coulomb-blockade electrometers},\ }}\href@noop {} {\bibfield  {journal}
  {\bibinfo  {journal} {IEEE Transactions on Applied Superconductivity}\
  }\textbf {\bibinfo {volume} {5}},\ \bibinfo {pages} {3085} (\bibinfo {year}
  {1995})}\BibitemShut {NoStop}%
\bibitem [{\citenamefont {Wolf}\ \emph {et~al.}(1997)\citenamefont {Wolf},
  \citenamefont {Ahlers}, \citenamefont {Niemeyer}, \citenamefont {Scherer},
  \citenamefont {Weimann}, \citenamefont {Zorin}, \citenamefont {Krupenin},
  \citenamefont {Lotkhov},\ and\ \citenamefont {Presnov}}]{Wolf1997}%
  \BibitemOpen
  \bibfield  {author} {\bibinfo {author} {\bibfnamefont {H.}~\bibnamefont
  {Wolf}}, \bibinfo {author} {\bibfnamefont {F.~J.}\ \bibnamefont {Ahlers}},
  \bibinfo {author} {\bibfnamefont {J.}~\bibnamefont {Niemeyer}}, \bibinfo
  {author} {\bibfnamefont {H.}~\bibnamefont {Scherer}}, \bibinfo {author}
  {\bibfnamefont {T.}~\bibnamefont {Weimann}}, \bibinfo {author} {\bibfnamefont
  {A.~B.}\ \bibnamefont {Zorin}}, \bibinfo {author} {\bibfnamefont {V.~A.}\
  \bibnamefont {Krupenin}}, \bibinfo {author} {\bibfnamefont {S.~V.}\
  \bibnamefont {Lotkhov}}, \ and\ \bibinfo {author} {\bibfnamefont {D.~E.}\
  \bibnamefont {Presnov}},\ }\bibfield  {title} {\emph {\bibinfo {title}
  {Investigation of the offset charge noise in single electron tunneling
  devices},\ }}\href@noop {} {\bibfield  {journal} {\bibinfo  {journal} {IEEE
  Transactions on Instrumentation and Measurement}\ }\textbf {\bibinfo {volume}
  {46}},\ \bibinfo {pages} {303} (\bibinfo {year} {1997})}\BibitemShut
  {NoStop}%
\bibitem [{\citenamefont {Kenyon}\ \emph {et~al.}(2000)\citenamefont {Kenyon},
  \citenamefont {Lobb},\ and\ \citenamefont {Wellstood}}]{Kenyon2000}%
  \BibitemOpen
  \bibfield  {author} {\bibinfo {author} {\bibfnamefont {M.}~\bibnamefont
  {Kenyon}}, \bibinfo {author} {\bibfnamefont {C.~J.}\ \bibnamefont {Lobb}}, \
  and\ \bibinfo {author} {\bibfnamefont {F.~C.}\ \bibnamefont {Wellstood}},\
  }\bibfield  {title} {\emph {\bibinfo {title} {Temperature dependence of
  low-frequency noise in al-al2o3-al single-electron transistors},\ }}\href
  {\doibase 10.1063/1.1312846} {\bibfield  {journal} {\bibinfo  {journal}
  {Journal of Applied Physics}\ }\textbf {\bibinfo {volume} {88}},\ \bibinfo
  {pages} {6536} (\bibinfo {year} {2000})}\BibitemShut {NoStop}%
\bibitem [{\citenamefont {Nakamura}\ \emph {et~al.}(2002)\citenamefont
  {Nakamura}, \citenamefont {Pashkin}, \citenamefont {Yamamoto},\ and\
  \citenamefont {Tsai}}]{Nakamura2002}%
  \BibitemOpen
  \bibfield  {author} {\bibinfo {author} {\bibfnamefont {Y.}~\bibnamefont
  {Nakamura}}, \bibinfo {author} {\bibfnamefont {Y.~A.}\ \bibnamefont
  {Pashkin}}, \bibinfo {author} {\bibfnamefont {T.}~\bibnamefont {Yamamoto}}, \
  and\ \bibinfo {author} {\bibfnamefont {J.~S.}\ \bibnamefont {Tsai}},\
  }\bibfield  {title} {\emph {\bibinfo {title} {Charge echo in a cooper-pair
  box},\ }}\href {https://link.aps.org/doi/10.1103/PhysRevLett.88.047901}
  {\bibfield  {journal} {\bibinfo  {journal} {Phys. Rev. Lett.}\ }\textbf
  {\bibinfo {volume} {88}},\ \bibinfo {pages} {047901} (\bibinfo {year}
  {2002})}\BibitemShut {NoStop}%
\bibitem [{\citenamefont {Gustafsson}\ \emph {et~al.}(2013)\citenamefont
  {Gustafsson}, \citenamefont {Pourkabirian}, \citenamefont {Johansson},
  \citenamefont {Clarke},\ and\ \citenamefont {Delsing}}]{Gustafsson2013}%
  \BibitemOpen
  \bibfield  {author} {\bibinfo {author} {\bibfnamefont {M.~V.}\ \bibnamefont
  {Gustafsson}}, \bibinfo {author} {\bibfnamefont {A.}~\bibnamefont
  {Pourkabirian}}, \bibinfo {author} {\bibfnamefont {G.}~\bibnamefont
  {Johansson}}, \bibinfo {author} {\bibfnamefont {J.}~\bibnamefont {Clarke}}, \
  and\ \bibinfo {author} {\bibfnamefont {P.}~\bibnamefont {Delsing}},\
  }\bibfield  {title} {\emph {\bibinfo {title} {Thermal properties of charge
  noise sources},\ }}\href
  {https://link.aps.org/doi/10.1103/PhysRevB.88.245410} {\bibfield  {journal}
  {\bibinfo  {journal} {PRB}\ }\textbf {\bibinfo {volume} {88}},\ \bibinfo
  {pages} {245410} (\bibinfo {year} {2013})}\BibitemShut {NoStop}%
\bibitem [{\citenamefont {Freeman}\ \emph {et~al.}(2016)\citenamefont
  {Freeman}, \citenamefont {Schoenfield},\ and\ \citenamefont
  {Jiang}}]{Freeman2016}%
  \BibitemOpen
  \bibfield  {author} {\bibinfo {author} {\bibfnamefont {B.~M.}\ \bibnamefont
  {Freeman}}, \bibinfo {author} {\bibfnamefont {J.~S.}\ \bibnamefont
  {Schoenfield}}, \ and\ \bibinfo {author} {\bibfnamefont {H.}~\bibnamefont
  {Jiang}},\ }\bibfield  {title} {\emph {\bibinfo {title} {Comparison of low
  frequency charge noise in identically patterned si/sio2 and si/sige quantum
  dots},\ }}\href {\doibase 10.1063/1.4954700} {\bibfield  {journal} {\bibinfo
  {journal} {Appl. Phys. Lett.}\ }\textbf {\bibinfo {volume} {108}},\ \bibinfo
  {pages} {253108} (\bibinfo {year} {2016})}\BibitemShut {NoStop}%
\bibitem [{\citenamefont {Dutta}\ and\ \citenamefont {Horn}(1981)}]{Dutta1981}%
  \BibitemOpen
  \bibfield  {author} {\bibinfo {author} {\bibfnamefont {P.}~\bibnamefont
  {Dutta}}\ and\ \bibinfo {author} {\bibfnamefont {P.~M.}\ \bibnamefont
  {Horn}},\ }\bibfield  {title} {\emph {\bibinfo {title} {Low-frequency
  fluctuations in solids: $\frac{1}{f}$ noise},\ }}\href
  {https://link.aps.org/doi/10.1103/RevModPhys.53.497} {\bibfield  {journal}
  {\bibinfo  {journal} {RMP}\ }\textbf {\bibinfo {volume} {53}},\ \bibinfo
  {pages} {497} (\bibinfo {year} {1981})}\BibitemShut {NoStop}%
\bibitem [{\citenamefont {Clemens~Müller}(2017)}]{Muller2017}%
  \BibitemOpen
  \bibfield  {author} {\bibinfo {author} {\bibfnamefont {J.~L.}\ \bibnamefont
  {Clemens~Müller}, \bibfnamefont {Jared H.~Cole}},\ }\bibfield  {title}
  {\emph {\bibinfo {title} {Towards understanding two-level-systems in
  amorphous solids - insights from quantum circuits},\ }}\href@noop {}
  {\bibfield  {journal} {\bibinfo  {journal} {arXiv:1705.01108
  [cond-mat.mes-hall]}\ } (\bibinfo {year} {2017})}\BibitemShut {NoStop}%
\bibitem [{\citenamefont {Rist{\`e}}\ \emph {et~al.}(2013)\citenamefont
  {Rist{\`e}}, \citenamefont {Bultink}, \citenamefont {Tiggelman},
  \citenamefont {Schouten}, \citenamefont {Lehnert},\ and\ \citenamefont
  {DiCarlo}}]{Riste2013}%
  \BibitemOpen
  \bibfield  {author} {\bibinfo {author} {\bibfnamefont {D.}~\bibnamefont
  {Rist{\`e}}}, \bibinfo {author} {\bibfnamefont {C.~C.}\ \bibnamefont
  {Bultink}}, \bibinfo {author} {\bibfnamefont {M.~J.}\ \bibnamefont
  {Tiggelman}}, \bibinfo {author} {\bibfnamefont {R.~N.}\ \bibnamefont
  {Schouten}}, \bibinfo {author} {\bibfnamefont {K.~W.}\ \bibnamefont
  {Lehnert}}, \ and\ \bibinfo {author} {\bibfnamefont {L.}~\bibnamefont
  {DiCarlo}},\ }\bibfield  {title} {\emph {\bibinfo {title} {Millisecond
  charge-parity fluctuations and induced decoherence in a superconducting
  transmon qubit},\ }}\href {https://doi.org/10.1038/ncomms2936} {\bibfield
  {journal} {\bibinfo  {journal} {Nature Communications}\ }\textbf {\bibinfo
  {volume} {4}},\ \bibinfo {pages} {1913} (\bibinfo {year} {2013})}\BibitemShut
  {NoStop}%
\bibitem [{\citenamefont {Serniak}\ \emph {et~al.}(2018)\citenamefont
  {Serniak}, \citenamefont {Hays}, \citenamefont {de~Lange}, \citenamefont
  {Diamond}, \citenamefont {Shankar}, \citenamefont {Burkhart}, \citenamefont
  {Frunzio}, \citenamefont {Houzet},\ and\ \citenamefont
  {Devoret}}]{Serniak2018}%
  \BibitemOpen
  \bibfield  {author} {\bibinfo {author} {\bibfnamefont {K.}~\bibnamefont
  {Serniak}}, \bibinfo {author} {\bibfnamefont {M.}~\bibnamefont {Hays}},
  \bibinfo {author} {\bibfnamefont {G.}~\bibnamefont {de~Lange}}, \bibinfo
  {author} {\bibfnamefont {S.}~\bibnamefont {Diamond}}, \bibinfo {author}
  {\bibfnamefont {S.}~\bibnamefont {Shankar}}, \bibinfo {author} {\bibfnamefont
  {L.~D.}\ \bibnamefont {Burkhart}}, \bibinfo {author} {\bibfnamefont
  {L.}~\bibnamefont {Frunzio}}, \bibinfo {author} {\bibfnamefont
  {M.}~\bibnamefont {Houzet}}, \ and\ \bibinfo {author} {\bibfnamefont {M.~H.}\
  \bibnamefont {Devoret}},\ }\bibfield  {title} {\emph {\bibinfo {title} {Hot
  nonequilibrium quasiparticles in transmon qubits},\ }}\href {\doibase
  10.1103/PhysRevLett.121.157701} {\bibfield  {journal} {\bibinfo  {journal}
  {Phys. Rev. Lett.}\ }\textbf {\bibinfo {volume} {121}},\ \bibinfo {pages}
  {157701} (\bibinfo {year} {2018})}\BibitemShut {NoStop}%
\bibitem [{\citenamefont {for~further information.}()}]{SuppMat}%
  \BibitemOpen
  \bibfield  {author} {\bibinfo {author} {\bibfnamefont {S.~S.~M.}\
  \bibnamefont {for~further information.}},\ }\href@noop {} {}\BibitemShut
  {NoStop}%
\bibitem [{\citenamefont {devices were originally designed to probe
  correlations of low-frequency flux}\ and\ \citenamefont {charge noise;~these
  measurements}()}]{ChargeFluxNote}%
  \BibitemOpen
  \bibfield  {author} {\bibinfo {author} {\bibfnamefont {T.}~\bibnamefont
  {devices were originally designed to probe correlations of low-frequency
  flux}}\ and\ \bibinfo {author} {\bibfnamefont {a.~d. i. t. S.~M.}\
  \bibnamefont {charge noise;~these measurements}, \bibfnamefont {which set an
  upper limit on flux-charge correlations at~6\%}},\ }\href@noop {}
  {}\BibitemShut {NoStop}%
\bibitem [{\citenamefont {Schreier}\ \emph {et~al.}(2008)\citenamefont
  {Schreier}, \citenamefont {Houck}, \citenamefont {Koch}, \citenamefont
  {Schuster}, \citenamefont {Johnson}, \citenamefont {Chow}, \citenamefont
  {Gambetta}, \citenamefont {Majer}, \citenamefont {Frunzio}, \citenamefont
  {Devoret}, \citenamefont {Girvin},\ and\ \citenamefont
  {Schoelkopf}}]{Schreier2008}%
  \BibitemOpen
  \bibfield  {author} {\bibinfo {author} {\bibfnamefont {J.~A.}\ \bibnamefont
  {Schreier}}, \bibinfo {author} {\bibfnamefont {A.~A.}\ \bibnamefont {Houck}},
  \bibinfo {author} {\bibfnamefont {J.}~\bibnamefont {Koch}}, \bibinfo {author}
  {\bibfnamefont {D.~I.}\ \bibnamefont {Schuster}}, \bibinfo {author}
  {\bibfnamefont {B.~R.}\ \bibnamefont {Johnson}}, \bibinfo {author}
  {\bibfnamefont {J.~M.}\ \bibnamefont {Chow}}, \bibinfo {author}
  {\bibfnamefont {J.~M.}\ \bibnamefont {Gambetta}}, \bibinfo {author}
  {\bibfnamefont {J.}~\bibnamefont {Majer}}, \bibinfo {author} {\bibfnamefont
  {L.}~\bibnamefont {Frunzio}}, \bibinfo {author} {\bibfnamefont {M.~H.}\
  \bibnamefont {Devoret}}, \bibinfo {author} {\bibfnamefont {S.~M.}\
  \bibnamefont {Girvin}}, \ and\ \bibinfo {author} {\bibfnamefont {R.~J.}\
  \bibnamefont {Schoelkopf}},\ }\bibfield  {title} {\emph {\bibinfo {title}
  {Suppressing charge noise decoherence in superconducting charge qubits},\
  }}\href {https://link.aps.org/doi/10.1103/PhysRevB.77.180502} {\bibfield
  {journal} {\bibinfo  {journal} {PRB}\ }\textbf {\bibinfo {volume} {77}},\
  \bibinfo {pages} {180502} (\bibinfo {year} {2008})}\BibitemShut {NoStop}%
\bibitem [{\citenamefont {Lutchyn}\ \emph {et~al.}(2006)\citenamefont
  {Lutchyn}, \citenamefont {Glazman},\ and\ \citenamefont
  {Larkin}}]{Lutchyn2006}%
  \BibitemOpen
  \bibfield  {author} {\bibinfo {author} {\bibfnamefont {R.~M.}\ \bibnamefont
  {Lutchyn}}, \bibinfo {author} {\bibfnamefont {L.~I.}\ \bibnamefont
  {Glazman}}, \ and\ \bibinfo {author} {\bibfnamefont {A.~I.}\ \bibnamefont
  {Larkin}},\ }\bibfield  {title} {\emph {\bibinfo {title} {Kinetics of the
  superconducting charge qubit in the presence of a quasiparticle},\ }}\href
  {https://link.aps.org/doi/10.1103/PhysRevB.74.064515} {\bibfield  {journal}
  {\bibinfo  {journal} {PRB}\ }\textbf {\bibinfo {volume} {74}},\ \bibinfo
  {pages} {064515} (\bibinfo {year} {2006})}\BibitemShut {NoStop}%
\bibitem [{\citenamefont {Yan}\ \emph {et~al.}(2012)\citenamefont {Yan},
  \citenamefont {Bylander}, \citenamefont {Gustavsson}, \citenamefont
  {Yoshihara}, \citenamefont {Harrabi}, \citenamefont {Cory}, \citenamefont
  {Orlando}, \citenamefont {Nakamura}, \citenamefont {Tsai},\ and\
  \citenamefont {Oliver}}]{Yan2012}%
  \BibitemOpen
  \bibfield  {author} {\bibinfo {author} {\bibfnamefont {F.}~\bibnamefont
  {Yan}}, \bibinfo {author} {\bibfnamefont {J.}~\bibnamefont {Bylander}},
  \bibinfo {author} {\bibfnamefont {S.}~\bibnamefont {Gustavsson}}, \bibinfo
  {author} {\bibfnamefont {F.}~\bibnamefont {Yoshihara}}, \bibinfo {author}
  {\bibfnamefont {K.}~\bibnamefont {Harrabi}}, \bibinfo {author} {\bibfnamefont
  {D.~G.}\ \bibnamefont {Cory}}, \bibinfo {author} {\bibfnamefont {T.~P.}\
  \bibnamefont {Orlando}}, \bibinfo {author} {\bibfnamefont {Y.}~\bibnamefont
  {Nakamura}}, \bibinfo {author} {\bibfnamefont {J.-S.}\ \bibnamefont {Tsai}},
  \ and\ \bibinfo {author} {\bibfnamefont {W.~D.}\ \bibnamefont {Oliver}},\
  }\bibfield  {title} {\emph {\bibinfo {title} {Spectroscopy of low-frequency
  noise and its temperature dependence in a superconducting qubit},\ }}\href
  {https://link.aps.org/doi/10.1103/PhysRevB.85.174521} {\bibfield  {journal}
  {\bibinfo  {journal} {PRB}\ }\textbf {\bibinfo {volume} {85}},\ \bibinfo
  {pages} {174521} (\bibinfo {year} {2012})}\BibitemShut {NoStop}%
\bibitem [{\citenamefont {Quintana}\ \emph {et~al.}(2017)\citenamefont
  {Quintana}, \citenamefont {Chen}, \citenamefont {Sank}, \citenamefont
  {Petukhov}, \citenamefont {White}, \citenamefont {Kafri}, \citenamefont
  {Chiaro}, \citenamefont {Megrant}, \citenamefont {Barends}, \citenamefont
  {Campbell}, \citenamefont {Chen}, \citenamefont {Dunsworth}, \citenamefont
  {Fowler}, \citenamefont {Graff}, \citenamefont {Jeffrey}, \citenamefont
  {Kelly}, \citenamefont {Lucero}, \citenamefont {Mutus}, \citenamefont
  {Neeley}, \citenamefont {Neill}, \citenamefont {O’Malley}, \citenamefont
  {Roushan}, \citenamefont {Shabani}, \citenamefont {Smelyanskiy},
  \citenamefont {Vainsencher}, \citenamefont {Wenner}, \citenamefont {Neven},\
  and\ \citenamefont {Martinis}}]{Quintana2017}%
  \BibitemOpen
  \bibfield  {author} {\bibinfo {author} {\bibfnamefont {C.~M.}\ \bibnamefont
  {Quintana}}, \bibinfo {author} {\bibfnamefont {Y.}~\bibnamefont {Chen}},
  \bibinfo {author} {\bibfnamefont {D.}~\bibnamefont {Sank}}, \bibinfo {author}
  {\bibfnamefont {A.~G.}\ \bibnamefont {Petukhov}}, \bibinfo {author}
  {\bibfnamefont {T.~C.}\ \bibnamefont {White}}, \bibinfo {author}
  {\bibfnamefont {D.}~\bibnamefont {Kafri}}, \bibinfo {author} {\bibfnamefont
  {B.}~\bibnamefont {Chiaro}}, \bibinfo {author} {\bibfnamefont
  {A.}~\bibnamefont {Megrant}}, \bibinfo {author} {\bibfnamefont
  {R.}~\bibnamefont {Barends}}, \bibinfo {author} {\bibfnamefont
  {B.}~\bibnamefont {Campbell}}, \bibinfo {author} {\bibfnamefont
  {Z.}~\bibnamefont {Chen}}, \bibinfo {author} {\bibfnamefont {A.}~\bibnamefont
  {Dunsworth}}, \bibinfo {author} {\bibfnamefont {A.~G.}\ \bibnamefont
  {Fowler}}, \bibinfo {author} {\bibfnamefont {R.}~\bibnamefont {Graff}},
  \bibinfo {author} {\bibfnamefont {E.}~\bibnamefont {Jeffrey}}, \bibinfo
  {author} {\bibfnamefont {J.}~\bibnamefont {Kelly}}, \bibinfo {author}
  {\bibfnamefont {E.}~\bibnamefont {Lucero}}, \bibinfo {author} {\bibfnamefont
  {J.~Y.}\ \bibnamefont {Mutus}}, \bibinfo {author} {\bibfnamefont
  {M.}~\bibnamefont {Neeley}}, \bibinfo {author} {\bibfnamefont
  {C.}~\bibnamefont {Neill}}, \bibinfo {author} {\bibfnamefont {P.~J.~J.}\
  \bibnamefont {O’Malley}}, \bibinfo {author} {\bibfnamefont
  {P.}~\bibnamefont {Roushan}}, \bibinfo {author} {\bibfnamefont
  {A.}~\bibnamefont {Shabani}}, \bibinfo {author} {\bibfnamefont {V.~N.}\
  \bibnamefont {Smelyanskiy}}, \bibinfo {author} {\bibfnamefont
  {A.}~\bibnamefont {Vainsencher}}, \bibinfo {author} {\bibfnamefont
  {J.}~\bibnamefont {Wenner}}, \bibinfo {author} {\bibfnamefont
  {H.}~\bibnamefont {Neven}}, \ and\ \bibinfo {author} {\bibfnamefont {J.~M.}\
  \bibnamefont {Martinis}},\ }\bibfield  {title} {\emph {\bibinfo {title}
  {Observation of classical-quantum crossover of $1/f$ flux noise and its
  paramagnetic temperature dependence},\ }}\href
  {https://link.aps.org/doi/10.1103/PhysRevLett.118.057702} {\bibfield
  {journal} {\bibinfo  {journal} {Phys. Rev. Lett.}\ }\textbf {\bibinfo
  {volume} {118}},\ \bibinfo {pages} {057702} (\bibinfo {year}
  {2017})}\BibitemShut {NoStop}%
\bibitem [{\citenamefont {Schoelkopf}\ \emph {et~al.}(1998)\citenamefont
  {Schoelkopf}, \citenamefont {Wahlgren}, \citenamefont {Kozhevnikov},
  \citenamefont {Delsing},\ and\ \citenamefont {Prober}}]{Schoelkopf1998}%
  \BibitemOpen
  \bibfield  {author} {\bibinfo {author} {\bibfnamefont {R.~J.}\ \bibnamefont
  {Schoelkopf}}, \bibinfo {author} {\bibfnamefont {P.}~\bibnamefont
  {Wahlgren}}, \bibinfo {author} {\bibfnamefont {A.~A.}\ \bibnamefont
  {Kozhevnikov}}, \bibinfo {author} {\bibfnamefont {P.}~\bibnamefont
  {Delsing}}, \ and\ \bibinfo {author} {\bibfnamefont {D.~E.}\ \bibnamefont
  {Prober}},\ }\bibfield  {title} {\emph {\bibinfo {title} {The radio-frequency
  single-electron transistor (rf-{S}{E}{T}): A fast and ultrasensitive
  electrometer},\ }}\href
  {http://science.sciencemag.org/content/280/5367/1238.abstract} {\bibfield
  {journal} {\bibinfo  {journal} {Science}\ }\textbf {\bibinfo {volume}
  {280}},\ \bibinfo {pages} {1238} (\bibinfo {year} {1998})}\BibitemShut
  {NoStop}%
\bibitem [{\citenamefont {Aassime}\ \emph {et~al.}(2001)\citenamefont
  {Aassime}, \citenamefont {Johansson}, \citenamefont {Wendin}, \citenamefont
  {Schoelkopf},\ and\ \citenamefont {Delsing}}]{Aassime2001}%
  \BibitemOpen
  \bibfield  {author} {\bibinfo {author} {\bibfnamefont {A.}~\bibnamefont
  {Aassime}}, \bibinfo {author} {\bibfnamefont {G.}~\bibnamefont {Johansson}},
  \bibinfo {author} {\bibfnamefont {G.}~\bibnamefont {Wendin}}, \bibinfo
  {author} {\bibfnamefont {R.~J.}\ \bibnamefont {Schoelkopf}}, \ and\ \bibinfo
  {author} {\bibfnamefont {P.}~\bibnamefont {Delsing}},\ }\bibfield  {title}
  {\emph {\bibinfo {title} {Radio-frequency single-electron transistor as
  readout device for qubits: Charge sensitivity and backaction},\ }}\href
  {https://link.aps.org/doi/10.1103/PhysRevLett.86.3376} {\bibfield  {journal}
  {\bibinfo  {journal} {Phys. Rev. Lett.}\ }\textbf {\bibinfo {volume} {86}},\
  \bibinfo {pages} {3376} (\bibinfo {year} {2001})}\BibitemShut {NoStop}%
\bibitem [{\citenamefont {Martinis}\ \emph {et~al.}(2005)\citenamefont
  {Martinis}, \citenamefont {Cooper}, \citenamefont {McDermott}, \citenamefont
  {Steffen}, \citenamefont {Ansmann}, \citenamefont {Osborn}, \citenamefont
  {Cicak}, \citenamefont {Oh}, \citenamefont {Pappas}, \citenamefont
  {Simmonds},\ and\ \citenamefont {Yu}}]{Martinis2005}%
  \BibitemOpen
  \bibfield  {author} {\bibinfo {author} {\bibfnamefont {J.~M.}\ \bibnamefont
  {Martinis}}, \bibinfo {author} {\bibfnamefont {K.~B.}\ \bibnamefont
  {Cooper}}, \bibinfo {author} {\bibfnamefont {R.}~\bibnamefont {McDermott}},
  \bibinfo {author} {\bibfnamefont {M.}~\bibnamefont {Steffen}}, \bibinfo
  {author} {\bibfnamefont {M.}~\bibnamefont {Ansmann}}, \bibinfo {author}
  {\bibfnamefont {K.~D.}\ \bibnamefont {Osborn}}, \bibinfo {author}
  {\bibfnamefont {K.}~\bibnamefont {Cicak}}, \bibinfo {author} {\bibfnamefont
  {S.}~\bibnamefont {Oh}}, \bibinfo {author} {\bibfnamefont {D.~P.}\
  \bibnamefont {Pappas}}, \bibinfo {author} {\bibfnamefont {R.~W.}\
  \bibnamefont {Simmonds}}, \ and\ \bibinfo {author} {\bibfnamefont {C.~C.}\
  \bibnamefont {Yu}},\ }\bibfield  {title} {\emph {\bibinfo {title}
  {Decoherence in josephson qubits from dielectric loss},\ }}\href
  {https://link.aps.org/doi/10.1103/PhysRevLett.95.210503} {\bibfield
  {journal} {\bibinfo  {journal} {Phys. Rev. Lett.}\ }\textbf {\bibinfo
  {volume} {95}},\ \bibinfo {pages} {210503} (\bibinfo {year}
  {2005})}\BibitemShut {NoStop}%
\bibitem [{\citenamefont {Gorodokin}\ and\ \citenamefont
  {Zemlyanov}(2004)}]{Gorodokin2004}%
  \BibitemOpen
  \bibfield  {author} {\bibinfo {author} {\bibfnamefont {V.}~\bibnamefont
  {Gorodokin}}\ and\ \bibinfo {author} {\bibfnamefont {D.}~\bibnamefont
  {Zemlyanov}},\ }\bibfield  {title} {\emph {\bibinfo {title} {Metallic
  contamination in silicon processing},\ }}in\ \href@noop {} {\emph {\bibinfo
  {booktitle} {2004 23rd IEEE Convention of Electrical and Electronics
  Engineers in Israel}}}\ (\bibinfo {year} {2004})\ pp.\ \bibinfo {pages}
  {157--160}\BibitemShut {NoStop}%
\bibitem [{\citenamefont {Constant}\ \emph {et~al.}(2000)\citenamefont
  {Constant}, \citenamefont {Tardif},\ and\ \citenamefont
  {Derrien}}]{Constant2000}%
  \BibitemOpen
  \bibfield  {author} {\bibinfo {author} {\bibfnamefont {I.}~\bibnamefont
  {Constant}}, \bibinfo {author} {\bibfnamefont {F.}~\bibnamefont {Tardif}}, \
  and\ \bibinfo {author} {\bibfnamefont {J.}~\bibnamefont {Derrien}},\
  }\bibfield  {title} {\emph {\bibinfo {title} {Deposition and removal of
  sodium contamination on silicon wafers},\ }}\href
  {http://dx.doi.org/10.1088/0268-1242/15/1/311} {\bibfield  {journal}
  {\bibinfo  {journal} {Semiconductor Science and Technology}\ }\textbf
  {\bibinfo {volume} {15}},\ \bibinfo {pages} {61} (\bibinfo {year}
  {2000})}\BibitemShut {NoStop}%
\bibitem [{\citenamefont {Turchette}\ \emph {et~al.}(2000)\citenamefont
  {Turchette}, \citenamefont {Kielpinski}, \citenamefont {King}, \citenamefont
  {Leibfried}, \citenamefont {Meekhof}, \citenamefont {Myatt}, \citenamefont
  {Rowe}, \citenamefont {Sackett}, \citenamefont {Wood}, \citenamefont {Itano},
  \citenamefont {Monroe},\ and\ \citenamefont {Wineland}}]{Turchette2000}%
  \BibitemOpen
  \bibfield  {author} {\bibinfo {author} {\bibfnamefont {Q.~A.}\ \bibnamefont
  {Turchette}}, \bibinfo {author} {\bibnamefont {Kielpinski}}, \bibinfo
  {author} {\bibfnamefont {B.~E.}\ \bibnamefont {King}}, \bibinfo {author}
  {\bibfnamefont {D.}~\bibnamefont {Leibfried}}, \bibinfo {author}
  {\bibfnamefont {D.~M.}\ \bibnamefont {Meekhof}}, \bibinfo {author}
  {\bibfnamefont {C.~J.}\ \bibnamefont {Myatt}}, \bibinfo {author}
  {\bibfnamefont {M.~A.}\ \bibnamefont {Rowe}}, \bibinfo {author}
  {\bibfnamefont {C.~A.}\ \bibnamefont {Sackett}}, \bibinfo {author}
  {\bibfnamefont {C.~S.}\ \bibnamefont {Wood}}, \bibinfo {author}
  {\bibfnamefont {W.~M.}\ \bibnamefont {Itano}}, \bibinfo {author}
  {\bibfnamefont {C.}~\bibnamefont {Monroe}}, \ and\ \bibinfo {author}
  {\bibfnamefont {D.~J.}\ \bibnamefont {Wineland}},\ }\bibfield  {title} {\emph
  {\bibinfo {title} {Heating of trapped ions from the quantum ground state},\
  }}\href {https://link.aps.org/doi/10.1103/PhysRevA.61.063418} {\bibfield
  {journal} {\bibinfo  {journal} {Phys. Rev. A}\ }\textbf {\bibinfo {volume}
  {61}},\ \bibinfo {pages} {063418} (\bibinfo {year} {2000})}\BibitemShut
  {NoStop}%
\bibitem [{\citenamefont {Deslauriers}\ \emph {et~al.}(2004)\citenamefont
  {Deslauriers}, \citenamefont {Haljan}, \citenamefont {Lee}, \citenamefont
  {Brickman}, \citenamefont {Blinov}, \citenamefont {Madsen},\ and\
  \citenamefont {Monroe}}]{Deslauriers2004}%
  \BibitemOpen
  \bibfield  {author} {\bibinfo {author} {\bibfnamefont {L.}~\bibnamefont
  {Deslauriers}}, \bibinfo {author} {\bibfnamefont {P.~C.}\ \bibnamefont
  {Haljan}}, \bibinfo {author} {\bibfnamefont {P.~J.}\ \bibnamefont {Lee}},
  \bibinfo {author} {\bibfnamefont {K.-A.}\ \bibnamefont {Brickman}}, \bibinfo
  {author} {\bibfnamefont {B.~B.}\ \bibnamefont {Blinov}}, \bibinfo {author}
  {\bibfnamefont {M.~J.}\ \bibnamefont {Madsen}}, \ and\ \bibinfo {author}
  {\bibfnamefont {C.}~\bibnamefont {Monroe}},\ }\bibfield  {title} {\emph
  {\bibinfo {title} {Zero-point cooling and low heating of trapped
  $^{111}\mathrm{Cd}^{+}$ ions},\ }}\href
  {https://link.aps.org/doi/10.1103/PhysRevA.70.043408} {\bibfield  {journal}
  {\bibinfo  {journal} {Phys. Rev. A}\ }\textbf {\bibinfo {volume} {70}},\
  \bibinfo {pages} {043408} (\bibinfo {year} {2004})}\BibitemShut {NoStop}%
\bibitem [{\citenamefont {Nolt}\ \emph {et~al.}(1985)\citenamefont {Nolt},
  \citenamefont {Radostitz}, \citenamefont {Carlotti}, \citenamefont {Carli},
  \citenamefont {Mencaraglia},\ and\ \citenamefont {Bonetti}}]{Nolt1985}%
  \BibitemOpen
  \bibfield  {author} {\bibinfo {author} {\bibfnamefont {I.~G.}\ \bibnamefont
  {Nolt}}, \bibinfo {author} {\bibfnamefont {J.~V.}\ \bibnamefont {Radostitz}},
  \bibinfo {author} {\bibfnamefont {M.}~\bibnamefont {Carlotti}}, \bibinfo
  {author} {\bibfnamefont {B.}~\bibnamefont {Carli}}, \bibinfo {author}
  {\bibfnamefont {F.}~\bibnamefont {Mencaraglia}}, \ and\ \bibinfo {author}
  {\bibfnamefont {A.}~\bibnamefont {Bonetti}},\ }\bibfield  {title} {\emph
  {\bibinfo {title} {Cosmic-ray backgrounds in infrared bolometers},\ }}\href
  {https://doi.org/10.1007/BF01011948} {\bibfield  {journal} {\bibinfo
  {journal} {International Journal of Infrared and Millimeter Waves}\ }\textbf
  {\bibinfo {volume} {6}},\ \bibinfo {pages} {707} (\bibinfo {year}
  {1985})}\BibitemShut {NoStop}%
\bibitem [{\citenamefont {Lanfranchi}\ \emph {et~al.}(1999)\citenamefont
  {Lanfranchi}, \citenamefont {Carli}, \citenamefont {Gignoli}, \citenamefont
  {Lee},\ and\ \citenamefont {Ridolfi}}]{Lanfranchi1999}%
  \BibitemOpen
  \bibfield  {author} {\bibinfo {author} {\bibfnamefont {M.}~\bibnamefont
  {Lanfranchi}}, \bibinfo {author} {\bibfnamefont {B.}~\bibnamefont {Carli}},
  \bibinfo {author} {\bibfnamefont {A.}~\bibnamefont {Gignoli}}, \bibinfo
  {author} {\bibfnamefont {C.}~\bibnamefont {Lee}}, \ and\ \bibinfo {author}
  {\bibfnamefont {M.}~\bibnamefont {Ridolfi}},\ }\bibfield  {title} {\emph
  {\bibinfo {title} {Cosmic-ray flux detected by an ir bolometer operated on
  board of a stratospheric aircraft},\ }}\href
  {http://www.sciencedirect.com/science/article/pii/S1350449599000249}
  {\bibfield  {journal} {\bibinfo  {journal} {Infrared Physics \& Technology}\
  }\textbf {\bibinfo {volume} {40}},\ \bibinfo {pages} {379} (\bibinfo {year}
  {1999})}\BibitemShut {NoStop}%
\bibitem [{\citenamefont {Karatsu}\ \emph {et~al.}(2019)\citenamefont
  {Karatsu}, \citenamefont {Endo}, \citenamefont {Bueno}, \citenamefont
  {de~Visser}, \citenamefont {Barends}, \citenamefont {Thoen}, \citenamefont
  {Murugesan}, \citenamefont {Tomita},\ and\ \citenamefont
  {Baselmans}}]{Karatsu2019}%
  \BibitemOpen
  \bibfield  {author} {\bibinfo {author} {\bibfnamefont {K.}~\bibnamefont
  {Karatsu}}, \bibinfo {author} {\bibfnamefont {A.}~\bibnamefont {Endo}},
  \bibinfo {author} {\bibfnamefont {J.}~\bibnamefont {Bueno}}, \bibinfo
  {author} {\bibfnamefont {P.~J.}\ \bibnamefont {de~Visser}}, \bibinfo {author}
  {\bibfnamefont {R.}~\bibnamefont {Barends}}, \bibinfo {author} {\bibfnamefont
  {D.~J.}\ \bibnamefont {Thoen}}, \bibinfo {author} {\bibfnamefont
  {V.}~\bibnamefont {Murugesan}}, \bibinfo {author} {\bibfnamefont
  {N.}~\bibnamefont {Tomita}}, \ and\ \bibinfo {author} {\bibfnamefont
  {J.~J.~A.}\ \bibnamefont {Baselmans}},\ }\bibfield  {title} {\emph {\bibinfo
  {title} {Mitigation of cosmic ray effect on microwave kinetic inductance
  detector arrays},\ }}\href {\doibase 10.1063/1.5052419} {\bibfield  {journal}
  {\bibinfo  {journal} {Appl. Phys. Lett.}\ }\textbf {\bibinfo {volume}
  {114}},\ \bibinfo {pages} {032601} (\bibinfo {year} {2019})}\BibitemShut
  {NoStop}%
\end{thebibliography}%

\pagebreak
\begin{center}
\textbf{\large Supplemental Materials: Anomalous Charge Noise in Superconducting Qubits}
\end{center}
\setcounter{equation}{0}
\setcounter{figure}{0}
\setcounter{table}{0}
\setcounter{page}{1}
\makeatletter
\renewcommand{\theequation}{S\arabic{equation}}
\renewcommand{\thefigure}{S\arabic{figure}}

\section{Fabrication details}

The devices are realized through single-layer fabrication on high-resistivity ($>20~\text{k}\Omega \cdot \text{cm}$) Si(100) wafers. After stripping the native SiO$_x$ with hydrofluoric acid, a 90~nm film of Nb is deposited at a rate of 45~nm/min. under conditions optimized to achieve slight compressive film stress.  All features except the qubit junctions are then defined with an i-line projection aligner, and the Nb is etched using a $\text{Cl}_2\text{/BCl}_3$ recipe in an inductively coupled plasma reactive ion etch tool.

Qubit junctions are then made using a standard Dolan bridge process with a MMA/PMMA stack.  
Following an \textit{in situ} ion mill to ensure good metallic contact with the base Nb layer, the $\text{Al-AlO}_{\text{x}}\text{-Al}$ junctions are formed by double-angle electron-beam evaporation of Al and thermal oxidation in a 90/10 Ar/O$_2$ mixture. 

\section{Comparison of $S_{\delta f}$ and $S_q$}

\begin{figure}[h!!]
\includegraphics[scale=0.41]{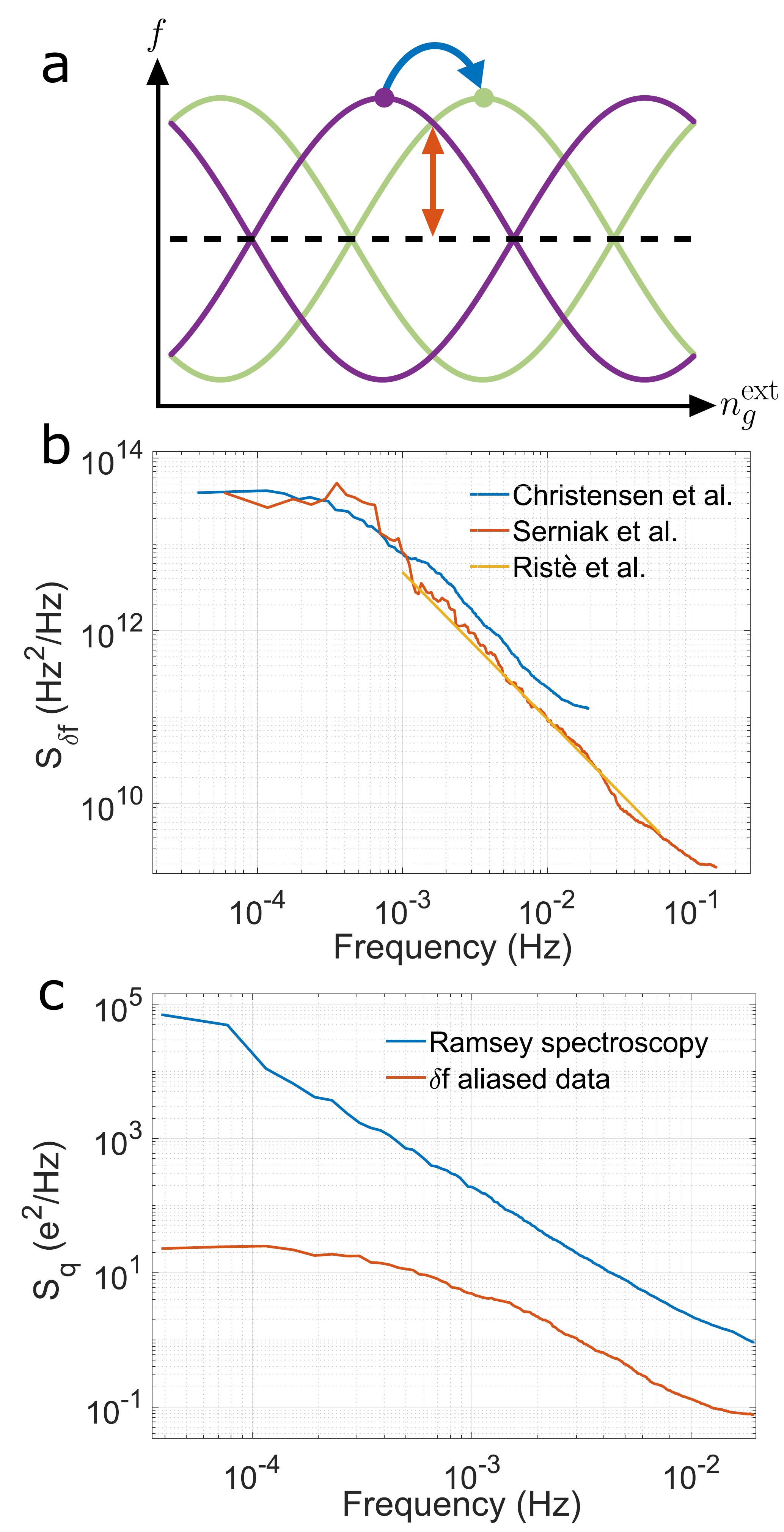}
\caption{\label{sfig:psd_comparison} Comparison of $S_{\delta f}$ and $S_q$. 
(a) Cartoon showing the two measurement techniques. The purple and green curves correspond to two instances of $\delta n_g$. The approach used in the main text (blue arrow) tracks the peak in the charge dispersion curve by sweeping $n_g^{\text{ext}}$ and fitting the resulting curve, yielding a dynamic range $[-0.5e,0.5e)$.  The technique in \cite{Riste2013,Serniak2018} tracks $\delta f$ (red arrow), which aliases all charge fluctuations into the interval $[0,0.5e]$.
(b) Plot of $S_{\delta f}$ for our data and \cite{Riste2013,Serniak2018}. The data sets show similar noise magnitude and exponent, suggesting a similar underlying noise mechanism. (c) Plot comparing $S_q$ taken with the two techniques.  The blue curve is the charge spectrum presented in the main text, while the orange curve is the result of mapping from offset charge to difference frequency and back to the offset charge interval $[0, 0.5e]$.}
\end{figure}

To compare with the power spectra of frequency fluctuations presented in \cite{Riste2013,Serniak2018}, we convert our time series of offset charge to time series of difference frequency using the relation $\delta f = |(\Delta\omega_{10}/2\pi) \cos 2 \pi n_g|$. The resulting power spectra of our data and of \cite{Riste2013,Serniak2018} are plotted in Fig.~\ref{sfig:psd_comparison}b.  The spectrum from the Rist{\`e} work is taken from the fit in Fig.~S4 of \cite{Riste2013}.  The spectrum from Serniak \textit{et al.} is calculated from a raw time series provided by those authors. As the difference frequency can only vary between 0 and $\Delta\omega_{10}/2\pi$, at long time scales there is a white noise ceiling, as the variation in difference frequency is capped. To allow direct comparison of the frequency noise, we normalize all frequency spectra to match the 600~kHz charge dispersion in our qubit A: the device of Serniak \textit{et al.} has a charge dispersion of 1.6~MHz, so we scale their spectrum by 0.14; while the device of Rist{\`e} \textit{et al.} has an 880~kHz charge dispersion, so we scale their spectrum by 0.46.

To extract the magnitude and exponent of the frequency noise, we fit the spectrum from $10^{-3}$ to $10^{-2}$~Hz for our data and from $10^{-3}$ to $10^{-1}$~Hz for \cite{Serniak2018}. We find for our measurements $\alpha = 1.76$ and $S_{\delta f}(\text{1\,Hz})= 5.9\times10^{7}~\text{Hz}^2/\text{Hz}$; for the data of Serniak \textit{et al.} we find $\alpha = 1.70$ and $S_{\delta f}(\text{1\,Hz})= 3.7\times10^{7}~\text{Hz}^2/\text{Hz}$; while Rist{\`e} \textit{et al}. cite $\alpha = 1.7$ and $S_{\delta f}(\text{1\,Hz})= 3.8\times10^{7}~\text{Hz}^2/\text{Hz}$.

As noted in the main text, the conversion from $\delta f$ to $n_g$ is not possible due to aliasing effects, thus there is no one-to-one mapping between the two functions (Fig.~\ref{sfig:psd_comparison}a). Our Ramsey-based approach can be thought of as tracking the full  spectroscopy curve as a function of $n_g$ by sweeping $n_g^{\text{ext}}$.  By tracking the full charge dispersion curve, we can monitor long-term drifts, with the only aliasing arising when two consecutive measurements differ in offset charge by a value larger than $\pm 0.5 e$. If instead (as in the works of \cite{Riste2013,Serniak2018}) no external bias is used and only the difference frequency at a single bias point is measured, then every data point is mapped to the interval $[0,0.5e]$.  An example of this is shown in Fig.~\ref{sfig:psd_comparison}a, where two different instances of $\delta n_g$ are displayed.  With access only to difference frequency at a single bias point, it is impossible to uniquely determine the value of $\delta n_g$ that produced the measured $\delta f$, and therefore an arbitrary choice must be made to assign a value to $\delta n_g$. This arbitrary choice must be made any time $n_g$ nears the boundary of the interval $[0,0.5e]$, as, e.g., $n_g = 0.6$ and $n_g = 0.4$ both correspond to the same difference frequency. Thus $\delta n_g$ can never leave the interval $[0,0.5e]$. Critically, this aliasing applies to \textit{every} measurement, whereas the aliasing from our approach only applies to \textit{changes} in offset charge between two measurements that exceed $\pm 0.5e$.

The impact of the aliasing is shown in Fig.~\ref{sfig:psd_comparison}c.  Here, we plot the charge noise power spectral density measured via Ramsey tomography in blue.  We then convert our data to a time series of $\delta f$, and finally map from $\delta f$ back to the gate charge interval $[0,0.5e]$. The power spectrum of the intentionally aliased data is plotted in orange. For the aliased data, we see a reduction of the noise exponent from 1.93 to 1.76, a reduction in the noise power by approximately two orders of magnitude, and a white noise ceiling at frequencies below $3 \times 10^{-4}$~Hz.

\section{Simulation of discrete charging events}

To generate the histogram in Fig.~\ref{fig:hist_and_cap}a, we perform a COMSOL simulation to calculate the induced charge on the qubit island associated with nucleation of a discrete $1e$ charge in the dielectric space between the qubit island and the circuit groundplane (Fig.~\ref{fig:hist_and_cap}b). We then permit both charge polarities, which would correspond either to the nucleation of both positive and negative charged particles, or to the adsorption/desorption of a single charged species.  We sample the entire groundplane cavity with uniform density, appropriately alias the data to account for the finite dynamic range of our charge measurement (so that, e.g., $0.6e$ is mapped to $-0.4e$), and histogram the results. This naive simulation yields surprising agreement with the measured distribution of discrete charge jumps. 

Within a picture of impingement of discrete charges in the dielectric cavity of the qubit groundplane, the measured rate of charge jumps corresponds to a flux of charged particles of $17/\text{cm}^2\cdot\text{s}$, which could be due, e.g., to a partial pressure of charged species of order $10^{-22}$~Torr. This pressure corresponds to roughly 0.4~ions in the Al box housing the sample, so the drift of charge might be better viewed as due to some element within the sample box that releases charge at a rate of $\sim 500$ particles per second. For example, the charge could be generated from the relaxation of strain in the PCB material used to couple signals into and out of the sample box or from the relaxation of strain in the dielectric substrate itself. 


\begin{figure}[h!!]
\includegraphics[width=\columnwidth]{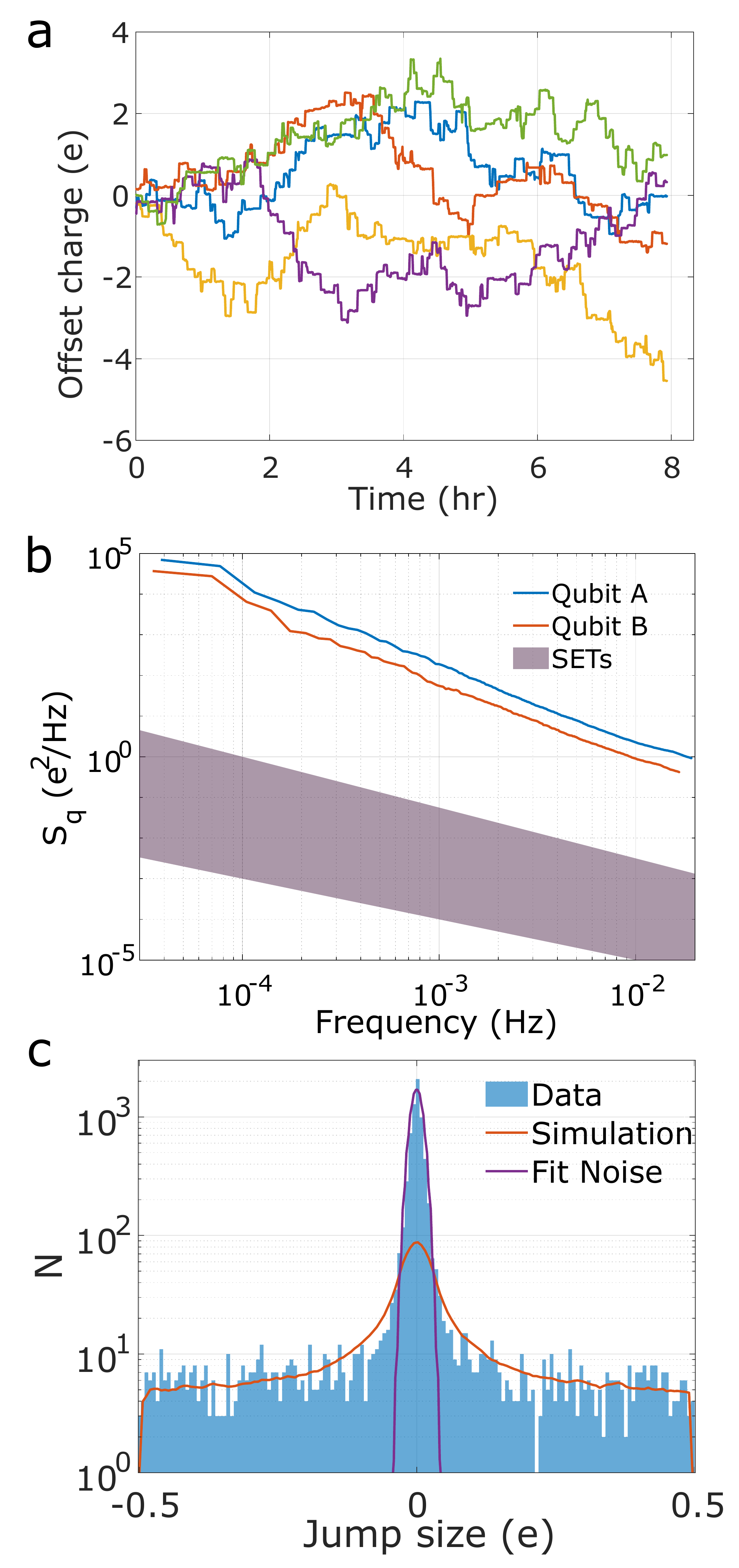}
\caption{\label{sfig:d2_stack} Measurements of qubit B. (a) Time series of offset charge in qubit B. (b) Low-frequency charge noise power spectral densities for qubits A and B, along with the range of charge noise seen in SETs. (c) Histogram of discrete charge jumps in qubit B. The solid traces represent the simulation of discrete charging events after the method of Section III (red) and Gaussian measurement uncertainty with width $0.01e$ (purple).}
\end{figure}

Alternatively, it could be that free charge is generated by cosmic rays that are absorbed in the qubit  substrate or in the material of the sample enclosure. However, the flux of cosmic rays is only $0.025 /\text{cm}^2\cdot\text{s}$ at sea level \cite{Nolt1985,Lanfranchi1999,Karatsu2019}, likely too low to account for observed rate of discrete charge jumps. 


\section{Measurements of qubit B}


A second, nominally identical device (qubit B) was characterized in the same cooldown that yielded the data presented in the main text. The only difference between the devices was a slightly reduced $E_J$ of 9.9~GHz for qubit B compared to 10.8~GHz for qubit A. Time series of fluctuating offset charge in qubit B are shown in Fig.~\ref{sfig:d2_stack}a.  The charge power spectral density calculated from this data shows $\alpha = 1.87$ and $S_q(\text{1\,Hz})= 1.6\times10^{-4}~\text{e}^2/\text{Hz}$ (Fig.~\ref{sfig:d2_stack}b). The histogram of discrete charge jumps measured in qubit B is shown in Fig.~\ref{sfig:d2_stack}c along with the results of simulations of the type described in the previous section.  In this case, the measurement cycle time was 30~s and the Gaussian central peak in the charge histogram corresponds to a fit uncertainty of 0.01$e$.  We find large discrete jumps in offset charge at a rate of one event per 290~s, corresponding to an impingement rate of $15/\text{cm}^2\cdot\text{s}$.

\section{Charge-Flux Noise Correlation}

\begin{figure*}[h]
\includegraphics[width=\textwidth]{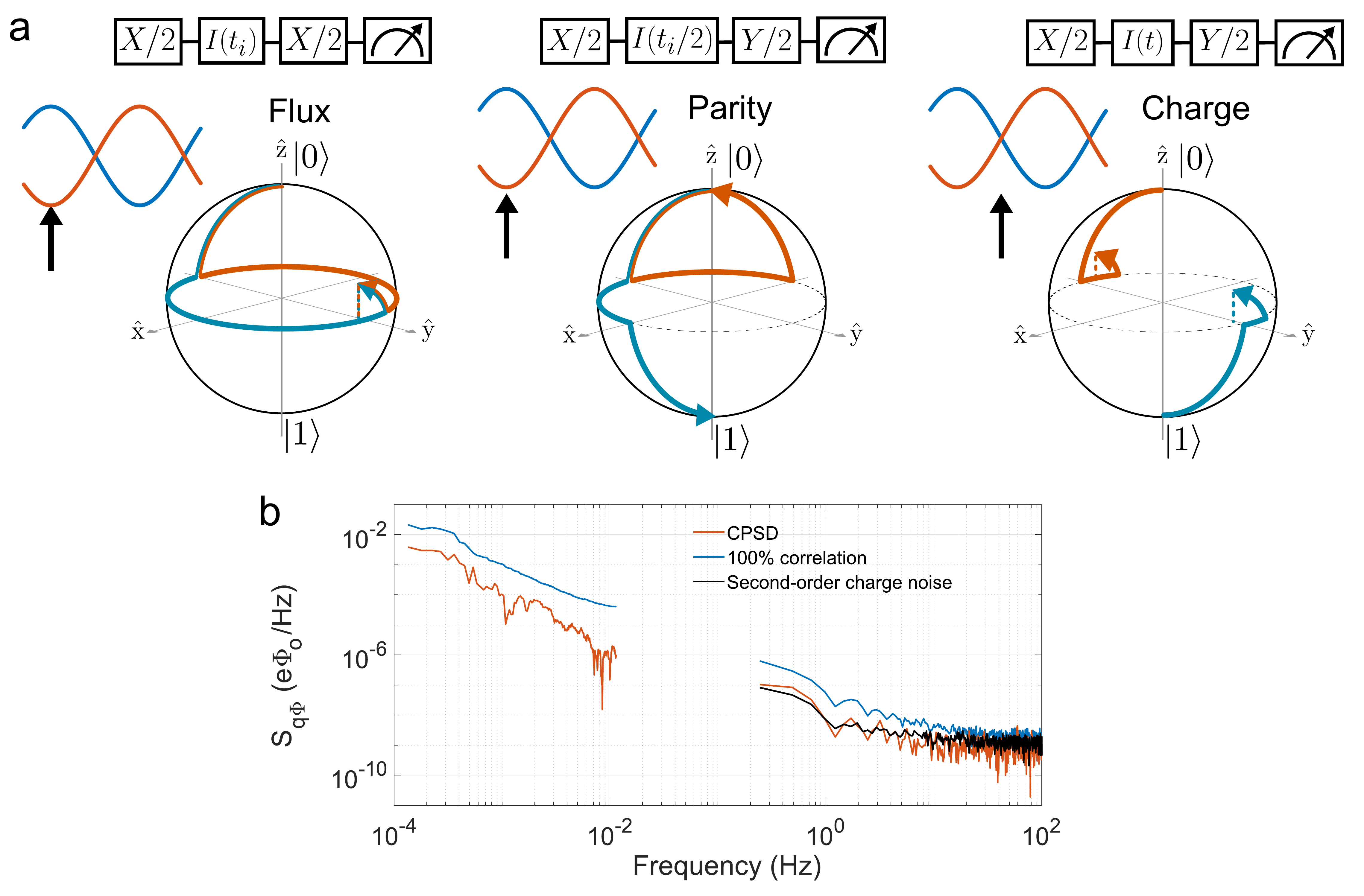}
\caption{\label{fig:FPC_sequence_CPSD} (a) Pulse sequence for the high-bandwidth measurement of the charge-flux CPSD $S_q\Phi$. The qubit is biased to a charge-insensitive, flux-sensitive point. A broadband $X/2$ gate followed by an appropriate idle time maps the two QP parity states to the same point on the equator of the Bloch sphere; a final $Y/2$ gate maps the accumulated phase to qubit population.  Following the flux measurement, we de-populate the readout cavity and immediately follow the measurement with the same QP parity / charge sequence described in the main text (Fig.~\ref{fig:pulseseq}a), with the charge noise measurement now conditioned on the flux noise measurement to account for the initial qubit state. (b) Cross power spectral density of charge and flux.  The blue curve is the geometric mean of the direct spectra $S_q$ and $S_\Phi$, the orange curve is the charge-flux CPSD $S_{q\Phi}$, and the black curve is the noise floor set by the CPSD between second-order charge noise and charge noise. The low-frequency spectra are derived from interleaved time series of fluctuating offset charge (from measurements of the type described in Fig.~\ref{fig:spectroscopy}) and offset flux (from a standard Ramsey experiment).  We average 250 spectra acquired at a rate of $5\times10^{-3}$ Hz, resulting in a noise floor of $S_{q\Phi}/\left(S_qS_\Phi\right)^{1/2} = 0.06$. The high-bandwidth measurement of charge-flux CPSD is dominated by second-order charge noise due to large random jumps in offset charge (black trace).}
\end{figure*}

A complete microscopic picture of low-frequency $1/f$ charge and flux noise is still lacking. To aid in the understanding of these noise sources, we performed a series of experiments on qubit A to probe charge-flux correlations, as observation of such correlations could be used to constrain possible microscopic models. 
Here we describe these measurements, which allow us to place an upper bound on the level of correlation between charge and flux fluctuators. 

The measurement protocol involves interleaved single-shot Ramsey sequences executed at appropriate bias points in flux and charge space that provide access to flux noise, QP parity, and charge noise. The high-bandwidth flux noise measurement is performed at a charge bias that yields zero first-order sensitivity to charge noise and maximal sensitivity to QP parity, as shown in Fig.~\ref{fig:FPC_sequence_CPSD}a. The idle time is fixed so that both parity bands are mapped to the same final qubit state. Since the experiment is conducted away from the upper flux sweet spot in order to achieve sensitivity to flux fluctuations, $E_J/E_C$ is reduced from its maximum value, resulting in a larger charge dispersion and a shorter idle time for the Ramsey-based flux noise measurements. By computing the cross power spectral density (CPSD) of single-shot flux and charge tomography scans, we probe correlations between flux and charge noise in the band from $2\times10^{-1}$~Hz to 100~Hz. 


A low-frequency cross spectrum is taken by interleaving the offset charge measurement with a low-bandwidth measurement of the flux.  This can be achieved by performing a standard Ramsey measurement at a flux-sensitive bias to determine the qubit free precession frequency and using the transfer function $d\omega_{10}/d\Phi$ to extract the fluctuating bias flux.  The Ramsey measurement is done with a fixed idle time (set by the frequency separation of the parity bands) and a varying phase of the final gate.  Fitting the resulting Ramsey measurement then gives an estimate of the flux deviation. From the separate charge and flux time series we compute the CPSD of charge and flux noise. 


In Fig.~\ref{fig:FPC_sequence_CPSD}b we combine the low- and high-bandwidth charge-flux CPSD in a single plot. To understand the strength of the correlation, it is informative to compare the cross spectrum $S_{q\Phi}$ with the geometric mean of the direct charge and flux spectra $\left(S_qS_\Phi\right)^{1/2}$. For maximally correlated noises, the magnitude of the CPSD will equal this quantity, with a well-defined phase across the spectrum.  Note, however, that two uncorrelated time series will also have a CPSD with a magnitude equal to the geometric mean of the direct spectra, although in this case the phases of the cross spectrum will be random.  As the magnitude of the sum of $N$ random phasors scales as $\sqrt{N}$, the noise floor $S_{q\Phi}/\left(S_qS_\Phi\right)^{1/2}$ is then set by $1/\sqrt{N}$, where $N$ is the number of averages per data point.  For the low-bandwidth measurement, we average 250 spectra at a data rate of $5\times10^{-3}$~Hz, corresponding to 36~hours of measurement, setting the noise floor $S_{q\Phi}/\left(S_qS_\Phi\right)^{1/2}$ at 0.06. Indeed, the upper bound on charge-flux correlation achieved around $10^{-2}$~Hz matches well with this noise floor set by finite averaging. 

The high-bandwidth measurement protocol involves a much higher repetition rate of 10~kHz, so that finite averaging is not an issue. However, in contrast to the low-bandwidth schemes that allow direct monitoring of charge and flux, the fast single-shot protocol monitors the accumulation of spurious phase, and we rely on known transfer functions to map this phase to fluctuating charge and flux. The flux measurement is executed at a point where we are first-order insensitive to charge noise (and \textit{vice versa}); however, the second-order sensitivity to charge noise at the charge sweet spot is non-negligible. Indeed, by performing a check experiment at a bias point that is insensitive to first order to both charge and flux noise, we find that the noise floor in our charge-flux CPSD is set by second-order charge noise (black trace in Fig.~\ref{fig:FPC_sequence_CPSD}b), putting an upper bound on charge-flux correlation at 1~Hz at the level of 0.1. 




\end{document}